\documentclass[10.7pt]{extarticle}
\usepackage{fullpage}
\usepackage[english]{babel}
\usepackage[utf8]{inputenc}
\usepackage{amsmath, amssymb, bm, tikz-cd}
\usepackage{graphicx}
\usepackage{subcaption}
\usepackage[colorinlistoftodos]{todonotes}
\usepackage{indentfirst}

\newcommand{\hiddenpower}[2] { \ifnum \numexpr#2=1 #1 \else #1^#2 \fi }
\numberwithin{equation}{section}

\newcommand{\dd}{\text{d}}
\newcommand{\pd}{\partial}

\usepackage{xparse}
\ExplSyntaxOn
\newcounter{diff_order}
\newcounter{diff_power}

\newcommand{\rawdiff}[3]
{
	\setcounter{diff_order}{0}
	\clist_map_inline:nn{#3}{\stepcounter{diff_order}}
	
	\frac{\hiddenpower{#1}{\thediff_order} #2}
	{
		\def\old_var{DefaultValue}
		\setcounter{diff_power}{0}
		
		\clist_map_inline:nn{#3}
		{
			\def\new_var{##1}
			\ifnum \thediff_power=0
				\stepcounter{diff_power}
			\else
				\tl_if_eq:NNTF \new_var \old_var
				{\stepcounter{diff_power}}
				{
					#1 \hiddenpower{\old_var}{\thediff_power}
					\setcounter{diff_power}{1}
				}
			\fi

			\def\old_var{##1}
		}
		
		#1 \hiddenpower{\old_var}{\thediff_power}
	}
}

\newcommand{\pdiff}[2]{\rawdiff{\pd}{#1}{#2}}

\ExplSyntaxOff

\newcommand{\lb}{\left(}
\newcommand{\rb}{\right)}

\renewcommand{\cos}[2][1]{\hiddenpower{\text{cos}}{#1} \lb #2 \rb}

\renewcommand{\sinh}[2][1]{\hiddenpower{\text{sinh}}{#1} \lb #2 \rb}
\renewcommand{\cosh}[2][1]{\hiddenpower{\text{cosh}}{#1} \lb #2 \rb}
\renewcommand{\tanh}[2][1]{\hiddenpower{\text{tanh}}{#1} \lb #2 \rb}

\newcommand{\sech}[2][1]{\hiddenpower{\text{sech}}{#1} \lb #2 \rb}

\renewcommand{\ln}[1]{\text{ln} \lb #1 \rb}
\newcommand{\e}[1]{\text{e}^{#1}}

\newcommand{\tr}[1]{\text{tr}\left\lbrace #1 \right\rbrace}

\newcommand{\comm}[2]{\ensuremath{\left[ #1, #2 \right]}}

\newcommand{\nbym}[2]{\ensuremath{#1{\times}#2}}
\newcommand{\nbyn}[1]{\nbym{#1}{#1}}

\author{}

\date{}

\begin{document}

\begin{center}
\strut\hfill


\vskip 0.45in

\noindent {{\bf{SOLITONS: CONSERVATION LAWS $\&$ DRESSING METHODS}}}\\
\vskip 0.3in

\noindent {\footnotesize { {{ANASTASIA DOIKOU \footnote{{\tt This paper is based on a series of lectures presented at Heriot-Watt University by A.D.}} AND IAIN FINDLAY}}}}
\vskip 0.4in

\noindent {\footnotesize School of Mathematical and Computer Sciences, Heriot-Watt University,\\
Edinburgh EH14 4AS, United Kingdom}


\vskip 0.1in

\noindent {\footnotesize {\tt E-mail: a.doikou@hw.ac.uk, iaf1@hw.ac.uk }}\\

\vskip 0.80in

\end{center}

\begin{abstract}

\noindent 
We review some of the fundamental notions associated to the theory of solitons. More precisely, we focus on the issue of conservation laws 
via the existence of the Lax pair and also on methods that provide solutions to partial or ordinary differential equations that
are associated to discrete or continuous integrable systems. The Riccati equation associated to a given continuous integrable system is also solved and 
hence suitable conserved quantities are derived. The notion of the Darboux-B\"acklund transformation is introduced and employed in order 
to obtain soliton solutions for specific examples of integrable equations. The Zakharov-Shabat dressing scheme and the Gelfand-Levitan-Marchenko 
equation are also introduced. Via this method generic solutions are produced, and integrable hierarchies are explicitly derived. Various discrete and 
continuous integrable models are  employed as  examples such as the Toda chain, the discrete non-linear Schr\"odinger model, the Korteweg-de Vries  
and non-linear Schr\"odinger equations as well as the sine-Gordon and Liouville models.
\end{abstract}

\date{}
\vskip 0.4in

\newpage

\tableofcontents

\newpage

\section{Introduction}
\label{sec:SW}

\noindent 
The main aim of this article is to introduce the key ideas and concepts on the theory of solitons and the associated partial differential equations (PDEs) or ordinary 
differential equations (ODEs). It is based on a series of lectures presented at Heriot-Watt University as a part of a graduate course, 
and is primarily addressed to graduate students or researchers who wish to acquire some fundamental knowledge on the theory of solitons.

Solitons are localized traveling wave packets that maintain their shape as they travel and emerge unaltered after interacting with each other.
Due to their ``particle-like'' nature they are of great mathematical significance with a plethora of realistic 
applications  including optical fibres, water wave tsunami formation, Bose-Einstein condensate and chemical reactions among others. Typical examples of equations that exhibit
solitonic solutions are the Korteweg-de Vries (KdV), the sinh-Gordon, and the non-linear Schr\"odinger (NLS) equations. All these are cases of 
exactly solvable equations known as {\it integrable} equations and can be described via a unifying scheme under the generic name of Inverse Scattering Transform \cite{IST} 
as will become transparent in these notes. Solitons were first documented by  Scott Russell in the $ 19^{\text{th}} $ century, who observed them along the union canal in 
Edinburgh (an aqueduct of which is now 
named after him, a mere 15 minute walk away from the Heriot-Watt campus) while investigating the most 
efficient design for canal boats. Despite this, it was not until much later that it was suggested as the solution to an equation, such as the 
shallow wave equation put forward by Korteweg and de Vries near the turn of the $ 20^{\text{th}} $ century.

As mentioned, solitons are closely related to the idea of integrability, a notion that  is associated to the existence of a large number of conserved quantities (i.e. symmetries), 
therefore a substantial  part of this presentation  is devoted to the systematic derivation of conserved quantities in integrable systems such as the Toda chain, the modified 
Korteweg-de Vries equation, the non-linear Schr\"odinger and sine-Gordon equations. In this context the notion of the Lax pair is introduced  \cite{Lax} together with the idea 
of the {\it monodromy} matrix as  well as the associated Riccati equation \cite{FT}, which lead to the derivation of the hierarchy of conserved charges. The main paradigm of a continuum system is the so called AKNS scheme \cite{AKNS}. This offers the main non-relativistic system, and it  is naturally associated to the NLS,
KdV and mKdV equations. Also, typical examples of relativistic systems such as the sine-Gordon and Liouville models are examined.
It  is worth noting that our description is exclusively based on the existence of a Lax pair. There is however the 
algebraic/Hamiltonian description (see for instance some typical textbooks and reviews \cite{FT}--\cite{doikou-classical} and references therein) 
based on the existence of the so-called classical $r$-matrix and an associated Poisson algebraic structure \cite{STS}. 
We shall not discuss this formulation in these notes, however we refer the interested reader to  \cite{FT}--\cite{doikou-classical}.

The other important issue we are addressing in these notes is the derivation of solutions of integrable PDEs/ODEs and the systematic construction of integrable hierarchies via the dressing process. 
There is a considerable amount of relevant literature on this well studied subject, however we are only referring to a limited number of relevant sources restricted to the main purposes 
of this article. The widely used dressing schemes are known under the names:  Darboux-B\"acklund dressing transformation, Zakharov-Shabat dressing,  Drinfeld-Sokolov method, 
and they are variations or rather formal descriptions of the inverse scattering transform (see e.g.  \cite{ZakharovShabat1}--\cite{STS2} as well as \cite{Darboux}--\cite{Kasman} 
and references therein). The dressing schemes not only provide a sound methodology of producing solutions of integrable  PDEs,  
but also allow the explicit construction of the Lax pairs of integrable hierarchies.  Note that in the Hamiltonian framework one can also construct the hierarchy 
of Lax pairs using a universal expression that requires the existence of the classical 
(or quantum) $r$-matrix \cite{STS, Korepin, DoikouFindlay}. 

We should mention that significant  issues, such as  Hirota's bilinear method or Sato's theory and connections to finite Grassmannians 
have  not been addressed in this presentation. We refer however the interested reader to  e.g. \cite{Hietarinta, Kodama, Jimbo, Drazin, Kasman} and references therein, 
for pedagogical discussions on these and related  issues.

\subsubsection*{The Korteweg-de Vries Equation}
\label{ssec:KdV}

\noindent 
Let us  now introduce one of the most important equations in the context of soliton theory that is the  Korteweg-de Vries equation.
The KdV equation  is a partial differential equation (PDE) that theoretically describes solitons, i.e. it admits solitonic solutions, and reads as
\begin{equation}
	{\partial u(x, t) \over \partial t} - 6u{\partial u(x,t) \over \partial x} + {\partial^3 u(x,t) \over \partial x^3} = 0. \label{eq:KdV}
\end{equation}

Let us derive below the one soliton solution based on rather simple considerations. Later in the text we shall introduce systematic ways of 
producing solitonic solutions based on the dressing method.
Inserting a travelling wave solution $ u(x, t) = f(\xi) $, $\xi = x -ct$  the KdV equation becomes:
\begin{equation}
	-cf' - 6ff' + f''' = 0 \nonumber
\end{equation}
where the prime denotes differentiation with respect to $\xi$, then recalling that $ (f^2)' = 2ff' $, this can be integrated to give:
\begin{equation}
	-cf - 3f^2 + f'' = A \nonumber
\end{equation}
\noindent for some arbitrary constant of integration $ A $. Multiplying through by $ f' $, this can be integrated again
(adding a further constant of integration $ B $):
\begin{equation}
	-\frac{c}{2}f^2 - f^3 + \frac{1}{2}(f')^2 = Af + B. \nonumber
\end{equation}

Choosing that $ f $ and $ f' $ both tend to 0 as $ \xi \to \pm \infty $ (i.e. that the traveling wave solution travels in one local 'lump'), both of the constants 
$ A $ and $ B $ will have to be set to 0, so this becomes:
\begin{equation}
	f' = \pm f \sqrt{2f + c}\ \Rightarrow\  \pm \int \dd \xi = \int \frac{1}{f \sqrt{2f + c}} d f. \nonumber
\end{equation}

To evaluate the right-hand integral, the substitution $ f = -\frac{c}{2} \sech[2]{\theta} $ is made, from which it follows that $ \dd f = c \tanh{\theta} \sech[2]{\theta} \dd \theta $. 
After this substitution, the integral becomes trivial (after recalling the hyperbolic trigonometric relation $ \tanh[2]{\theta} = 1 - \sech[2]{\theta} $):
\begin{align}
	\pm \xi - x_0 &= -\frac{2}{\sqrt{c}} \int d \theta \nonumber \\
	&= -\frac{2}{\sqrt{c}} \theta \nonumber
\end{align}

\noindent for some constant of integration $ x_0 $. Inserting this back into the substitution made earlier finally gives the solitonic solution to the KdV equation
\begin{equation}
	f(x - ct) = -\frac{c}{2} \sech[2]{\frac{\sqrt{c}}{2} \lb \pm (x - ct) - x_0 \rb}. \label{eq:KdVSoliton}
\end{equation}

\subsubsection*{The modified KdV equation}
\label{ssec:ISPKdV}

\noindent 
We are going to derive yet another significant integrable PDE i.e. the modified KdV (mKdV) equation, starting from the KdV equation.
The connection with the time independent Schr\"odinger equation will also be established. 

First, recall the KdV equation
\begin{equation}
	\pdiff{u}{t} - 6u\pdiff{u}{x} + \pdiff{u}{x, x, x} = 0. \nonumber
\end{equation}
We now introduce the so-called {\it Miura transformation}: $ u = v^2 + \pdiff{v}{x} $, and substituting this into the KdV equation we obtain
\begin{align}
	0 &= \pdiff{}{t} \lb v^2 + \pdiff{v}{x} \rb - 6\lb v^2 + \pdiff{v}{x} \rb \pdiff{}{x} \lb v^2 + \pdiff{v}{x} \rb + \pdiff{}{x, x, x} \lb v^2 + \pdiff{v}{x} 
\rb =\ldots\nonumber \\
	&= \lb 2v + \pdiff{}{x} \rb \lb \pdiff{v}{t} - 6v^2 \pdiff{v}{x} + \pdiff{v}{x, x, x} \rb \nonumber
\end{align}
and consequently $ v $ must obey an equation called the modified KdV equation
\begin{equation}
	\pdiff{v}{t} - 6v^2 \pdiff{v}{x} + \pdiff{v}{x, x, x} = 0. \label{eq:mKdV}
\end{equation}
We also introduce the field $ \psi $ such that $ v = \psi^{-1} \pdiff{\psi}{x} $, then the Miura transform becomes
\begin{align}
	u(x, t) &= \lb \psi^{-1} \pdiff{\psi}{x} \rb^2 + \pdiff{}{x} \lb \psi^{-1} \pdiff{\psi}{x} \rb= \psi^{-1} \pdiff{\psi}{x, x} \nonumber \\
	\Rightarrow 0 &= \pdiff{\psi}{x, x} - u \psi. \label{eq:KdV_SLFormJustu}
\end{align}

The latter expression has the form of the time independent Schr\"odinger equation with the potential given by the solution to 
the original KdV equation, $ u(x, t) $, e.g. the one-soliton solution.
To generalize this to non-zero spectrum $ \lambda $ of the Schr\"odinger equation the symmetries of the KdV equation need to be exploited, 
namely, the solution $u$ is invariant under 
the transformation $ u(x, t) \to \lambda + u(x + 6t\lambda, t) $, for some real number $ \lambda $. Then, 
this transformation introduces a factor $ \lambda \psi $:
\begin{equation}
	\pdiff{\psi}{x, x} + (\lambda - u) \psi = 0. \label{eq:KdV_SLForm}
\end{equation}

We shall discuss in what follows the systematic construction and solutions of integrable PDEs and ODEs, 
as well as the existence of the associated conserved quantities.

\section{The Lax Formulation}
\label{sec:Lax}

\noindent 
As already pointed out we are interested in integrable systems,  i.e. systems that possess a large number of conserved quantities,
can in principle be exactly solved, and usually display solitonic solutions. The issue of conservation laws in relation to such equations will be 
thoroughly discussed in the subsequent sections. 

The idea of integrability is synonymous to the existence of a pair of operators called the {\it Lax pair} \cite{Lax}, that are associated to a 
set of linear relations that comprise 
the so-called auxiliary linear problem. Consider in general a pair of operators $\big (L,\ M\big )$, the associated spectral problem 
and the evolution problem respectively:
\begin{eqnarray}
&& L\Psi = \zeta \Psi \nonumber\\
&& \dot \Psi = M \Psi, \label{A1}
\end{eqnarray}
where the ``dot'' denotes differentiation with respect to time.
Consistency of the two equations above,  provided also that $\dot \zeta =0$, yields the {\it zero curvature condition}:
\begin{equation}
\dot  L = \big [ M,\  L\big ], \label{zero1}
\end{equation}
where  we define the commutator $ \big [ X, Y \big] = XY - YX $. Expression (\ref{zero1}) provides the equations of motion of the system described by the given Lax pair. 
These equations are in general integrable PDEs or ODEs depending on the type of system under consideration, i.e. discrete or continuous. In general, the pair $(L,\  M)$ 
can be differential operators or  $d \times d$ matrices,  that depend on some classical fields,  solutions of the associated non-linear ODEs or PDEs. 
We examine below two fundamental  examples of such models i.e. the KdV equation and the Toda chain. In the subsequent sections we employ the 
notion of the Lax pair and the associated auxiliary linear problem (\ref{A1}) to construct continuous and discrete integrable models, and also identify 
the associated non-linear PDEs or ODEs.

\begin{itemize}

\item {\bf Differential  operators: the KdV equation}\\
Let us consider the generic pair of differential operators:
\begin{eqnarray}
	&&L(x,t)= -\partial_x^2 + u(x, t) \nonumber\\
&&M(x,t) = a \partial_x^3  +f(x,t)\partial_x + g(x,t)
 \label{eq:LaxSL}
\end{eqnarray}
where we use the notation $\partial_t f = {\partial f \over \partial t}$, $\partial_x^n f= {\partial^n f \over \partial x^n}$.
Assume also that the pair above satisfies the zero curvature condition. We can then identify the functions $f,\ g$ as well as the associated non-linear PDE:  
\begin{align}
	\pdiff{u}{t} \psi &= \left[ \lb a\pdiff{}{x, x, x} + f \pdiff{}{x} + g \rb \lb u - \pdiff{}{x, x} \rb - \lb u - \pdiff{}{x, x} \rb 
\lb a\pdiff{}{x, x, x} + f \pdiff{}{x} + g \rb \right] \psi \nonumber \\
	&= \lb a\pdiff{u}{x, x, x} + f \pdiff{u}{x} + \pdiff{g}{x, x} \rb \psi+ \lb 3a\pdiff{u}{x, x} + \pdiff{f}{x, x} + 2\pdiff{g}{x} 
\rb \pdiff{\psi}{x}  \nonumber\\
&+ \lb 3a\pdiff{u}{x} + 2\pdiff{f}{x} \rb \pdiff{\psi}{x, x} \nonumber
\end{align}
\noindent so if we want there to be no terms depending on derivatives of $ \psi $, then this gives us the definitions of 
$ f(x, t) $ and $ g(x, t) $, up to some constants of integration:
\begin{equation}
f(x, t) = -\frac{3}{2} a u,  ~~~~g(x, t) = -\frac{3}{4} a \partial_x u.
\label{eqs:LaxKdV_fg}
\end{equation}

These can then be inserted back into the equation for the time evolution of $ u $ to give the KdV equation (setting $ a = -4 $):
\begin{equation}
	\pdiff{u}{t} = 6u\pdiff{u}{x} - \pdiff{u}{x, x, x}. \label{eq:LaxKdV}
\end{equation}

\item {\bf $N \times N$ matrices: the Toda chain}\\
The Lax pair associated to the Toda chain \cite{Toda}  is given by (time dependence is always implied even when it is not explicitly stated)
\begin{eqnarray}
	&& L = \sum_{j = 1}^{N} p_j e_{jj} + \sum_{j = 1}^{N - 1} \e{(q_{j + 1} - q_j) / 2} \lb e_{j j + 1} + e_{j + 1,j} \rb \nonumber\\
&&  M =  {1\over 2}\sum_{j = 1}^{N - 1} \e{(q_{j + 1} - q_j) / 2} \lb e_{j j+1} -  e_{j+1 j }  \rb,
 \label{TodaL}
\end{eqnarray}
where $ e_{i j} $ is the $\nbyn{N}$ matrix with elements $(e_{ij})_{kl}$. Provided that the Lax pair satisfies the zero curvature condition (\ref{zero1}) we conclude
\begin{eqnarray}
	&& \dot q_j = p_j \nonumber \\
	&& \dot p_j  = \e{q_{j + 1} - q_j} - \e{q_j - q_{j - 1}}. \nonumber
\end{eqnarray}
Combining the two equations above we obtain the equations of motion for the Toda chain
\begin{equation}
	\ddot q_j = \e{q_{j + 1} - q_j} - \e{q_j - q_{j - 1}}. \label{eq:ALP_ExTodaqEvo}
\end{equation}

A solution of this equation is the discrete soliton expressed as
\begin{equation}
	q_j(t) = q_{+} + \ln{\frac{1 + \frac{\gamma}{1 - \e{-2\kappa}} \e{-2\kappa (j + 1) + \sigma \sinh{\kappa} t}}
{1 + \frac{\gamma}{1 - \e{-2\kappa}} \e{-2\kappa j + \sigma \sinh{\kappa} t}}} \label{eq:ALP_ExTodaSoliton}
\end{equation}
 where $ q_{+} $, $ \kappa $, $ \sigma $, and $ \gamma $ are constants, with $ \sigma = \pm 2 $. This can be verified by inspection.

\end{itemize}

\subsection{Conservation laws}
\label{ssec:ALP_Lax}
\noindent 
As mentioned integrability is associated to the existence of a large number of conservation laws, usually as many as the degrees of freedom of the system under consideration.
Based on the existence of a Lax pair and the relevant zero curvature condition we shall be able in what follows to systematically derive the conserved 
charges for various examples of integrable models such as the KdV equation, the Toda chain, the discrete NLS model and the AKNS system.

\subsubsection{The KdV equation}

\noindent  Recall the Miura transformation, introduced in the previous subsection
\begin{equation}
 u = v^2 + \pdiff{v}{x}, \label{Miura}
\end{equation}
and yet another transformation, called the Gardner transformation:
\begin{equation}
 v = \frac{1}{2\epsilon} + \epsilon w \label{Gardner}
\end{equation}
for some $ \epsilon $. Combining the two transformations (\ref{Miura}), (\ref{Gardner}) we conclude
\begin{equation}
	u = \frac{1}{4\epsilon^2} + w + \epsilon^2 w^2 + \epsilon \pdiff{w}{x}. \label{eq:Gardner}
\end{equation}
The additive constant in the latter expression can be removed by utilizing the symmetry of the KdV equation under the transformation $ u(x, t) \to u(x - 6\lambda t, t) - 
\lambda $, with $ \lambda = \frac{1}{4\epsilon^2} $. Then, inserting this into the KdV equation:
\begin{align}
	0 &= \pdiff{}{t} \lb w + \epsilon^2 w^2 + \epsilon \pdiff{w}{x} \rb - 6\lb w + \epsilon^2 w^2 + \epsilon \pdiff{w}{x} \rb \pdiff{}{x} \lb w + \epsilon^2 w^2 
+ \epsilon \pdiff{w}{x} \rb  \nonumber\\ 
& + \pdiff{}{x, x, x} \lb w  + \epsilon^2 w^2 + \epsilon \pdiff{w}{x} \rb  \\
	&= A + \epsilon \pdiff{A}{x} + 2\epsilon^2 wA, \nonumber
\end{align}
where we define  $ A = \pdiff{w}{t} - 6w\pdiff{w}{x} + \pdiff{w}{x, x, x} - 6\epsilon^2 w^2 \pdiff{w}{x} $, 
after some tedious but nevertheless straightforward computations.
If $ A = 0 $, then the original KdV equation is automatically satisfied. Thus, the KdV equation is equivalent to
\begin{equation}
	0 = \pdiff{w}{t} - 6\lb w + \epsilon^2 w^2 \rb \pdiff{w}{x} + \pdiff{w}{x, x, x}. \label{eq:KdV_Gardner}
\end{equation}

In the limit $ \epsilon \to 0 $, $ w \to u $,  the field $ w $ can be expressed as a formal  series expansion expansion about $ \epsilon \to 0 $:
\begin{equation}
w(x, t; \epsilon) = \sum_{n = 0}^{\infty} \epsilon^n w_n(x, t). \label{eq:GardnerExpansion}
\end{equation}

Choosing $ T = w $ and $ X = -3w^2 - 2\epsilon^2 w^3 + \pdiff{w}{x, x} $ it is easy to see that Eq. \eqref{eq:KdV_Gardner} reduces to $\partial_t T +\partial_xX =0$, 
and hence we conclude that $ \int_{-\infty}^{\infty} w \dd x $  is a constant of motion, provided that $w$ vanishes at $x\to \pm \infty$. 
Then, inserting the asymptotic expansion and considering each power of $ \epsilon $ in turn, it follows that $ \int_{-\infty}^{\infty} w_n \dd x $
is also a constant of motion for each $ n $. Therefore, there are actually an infinite number of constants of motion for the KdV equation!

As $ w $ must agree with $ u $ when $ \epsilon = 0 $,  $ w_0 $ can be immediately recognized as $ w_0 = u $. Then, the remaining $ w_n $ can be read by 
inserting this expansion \eqref{eq:GardnerExpansion} into the transformation \eqref{eq:Gardner}, giving a recursion relation (after removing the $ n = 0 $ case):
\begin{equation}
	0 = w_n + \sum_{\substack{i, j = 0 \\ i + j = n - 2}}^{\infty} w_i w_j + \pdiff{w_{n - 1}}{x}. \label{eq:KdV_CoMRecursion}
\end{equation}

The first four members of the series expansion read as, $w_0 = u$ and 
\begin{eqnarray}
	&& w_1 = -\pdiff{w_0}{x} = -\pdiff{u}{x} \nonumber\\
	&& w_2 = -w_0^2 - \pdiff{w_1}{x} = \pdiff{u}{x, x} - u^2 \nonumber\\
	&& w_3 = -2w_0 w_1 - \pdiff{w_2}{x} = 4u\pdiff{u}{x} - \pdiff{u}{x, x, x}. \nonumber
\end{eqnarray}

\subsubsection{Conserved quantities from $L$: The Toda chain}

\noindent 
We shall now be working with the general Lax pair $\big  (L,\ M \big ) $, where $ L $ and $ M $ are \nbyn{N} matrices that depend on some 
classical fields (solutions of the associated integrable ODEs/PDEs), as well as some spectral parameter $ \lambda \in \mathbb{C} $.
For this pair, recall that the zero-curvature condition is $ \dot L= \comm{M}{L} $, while the associated spectral problem is $ L \psi = \zeta \psi $. 
We introduce a function $ {\mathfrak g}^{(n)} $, defined to be the trace of the $ n^{th} $ power of $ L $, i.e. 
$ {\mathfrak g}^{(n)}(\lambda) = tr\big (L^n(\lambda) \big)$. 
The differential of this with respect to time can be easily calculated, using the properties of the trace 
(namely, that it pulls through differentiation and that it is cyclic):
\begin{eqnarray}
	\dot {\mathfrak g}^{(n)} &=& n\ tr\big( L^{n - 1} \dot  L  \big ) \nonumber \\
	&=& n\ tr \big ( L^{n - 1} (LM - ML) \big ) \nonumber \\
	&=&  n\ tr\big ( L^n M \big ) - n\ tr\big ( L^{n - 1} ML\big ) \nonumber\\
	&=& n\ tr \big ( L^n M \big ) - n\ tr\big (L^n M \big ) = 0. \nonumber
\end{eqnarray}
So for any value of $ n $, $ {\mathfrak g}^{(n)} $ is a conserved quantity.

To see how this can be used to yield the conserved quantities, we look at our typical example of an integrable model 
with discrete degrees of freedom: the Toda chain (\ref{TodaL}).  To find the conserved quantities $ t^{(n)} $, we start by taking the trace of $ L $:
\begin{equation}
{\mathfrak g}^{(1)}= tr(L)= \sum_{j = 1}^{N} p_j,  \label{eq:ALP_ExTodaP}
\end{equation}
which is the total momentum of the system. The second conserved quantity can be similarly found
\begin{align}
{\mathfrak g}^{(2)} &= tr\big ( L^2 \big ) =  \sum_{j = 1}^{N} p_j^2 + 2\sum_{j = 1}^{N - 1} \e{q_{j + 1} - q_j} \nonumber
\end{align}
\noindent and after multiplying by a factor of $ \frac{1}{2} $, this gives the Hamiltonian for this model
\begin{equation}
H = \frac{1}{2} \sum_{j = 1}^{N} p_j^2 + \sum_{j = 1}^{N-1} \e{q_{j + 1} - q_j}.\label{eq:ALP_ExTodaH}
\end{equation}

\subsection{Semi-discrete Lax-pairs $\&$ the auxiliary linear problem}
\label{ssec:DsALP}

\noindent 
We may now generalize the notion of the Lax pair to describe one dimensional discrete space and continuum time integrable systems. Typical examples of such  systems  are the
Toda  chain (which shall be revisited in this context) or the discrete version of the NLS model among others.
In this case the Lax pair $\big ( L,\ A\big )$ satisfies the semi-discrete auxiliary linear problem, expressed as (see e.g. \cite{FT})
\begin{eqnarray}
	&&\Psi_{j + 1}(t, \lambda) = L_j(t, \lambda) \Psi_j(t, \lambda)\nonumber \\
	&&\dot \Psi_j(t, \lambda)= A_j(t, \lambda) \Psi_j(t, \lambda). \label{eq:ALP_ALPA}
\end{eqnarray}
The Lax pair are $d \times d$ matrices depending on some fields and the spectral parameter. Here we shall only consider examples where $L,\ A$ are $2 \times 2$ matrices.

Differentiating the first of the equations (\ref{eq:ALP_ALPA}) with respect to time produces the semi-discrete zero curvature condition
\begin{align}
	& \dot \Psi_{j + 1} = \dot  L_j \Psi_j + L_j \dot \Psi_j\nonumber \\
	& A_{j + 1} L_j \Psi_j = \dot L_j\Psi_j + L_j A_j \Psi_j \nonumber
\end{align}
which gives:
\begin{equation}
	\dot L_j = A_{j + 1} L_j - L_j A_j.  \label{eq:ALP_dZCC}
\end{equation}

Let us also introduce the discrete space {\it monodromy matrix}:
\begin{equation}
 T_N(\lambda) = L_{N}(\lambda) \ldots L_1(\lambda), \label{mono1}
\end{equation}
which is a solution of the first equation of the auxiliary linear problem (\ref{eq:ALP_ALPA}). Let us also define the so-called 
{\it transfer matrix}: ${\mathfrak t}(\lambda) = tr \big (T(\lambda)\big ) $, 
then we can show using the discrete zero curvature condition that this is  constant in time. Indeed, consider the time derivatives of $ {\mathfrak t} $ 
(we suppress the subscript $N$ for brevity):
\begin{align}
	\dot {\mathfrak t}(\lambda)&= tr\big (\dot T(\lambda) \big )\nonumber \\
	&= tr\Big ( \sum_{j = 1}^{N} L_N(\lambda) \ldots  \dot L_j L_{j - 1} (\lambda)... L_1(\lambda) \Big )\nonumber \\
	&= \sum_{j = 1}^{N} tr\Big(L_N(\lambda) \ldots L_{j + 1}(\lambda) A_{j + 1}(\lambda) L_j (\lambda) \ldots L_1(\lambda)\Big) \nonumber\\
&  - \sum_{j = 1}^{N} tr\Big (L_N(\lambda) \ldots  L_j(\lambda) A_j (\lambda)L_{j - 1}(\lambda) \ldots L_1(\lambda)\Big )   \nonumber\\
&=  tr\Big (A_{N+1}(\lambda) T(\lambda) -T(\lambda) A_1(\lambda) \Big )=0, \nonumber
\end{align}
provided that periodic boundary conditions are imposed, i.e. $A_{N+1}(\lambda) = A_1(\lambda)$. 
The latter expression shows that $ {\mathfrak t}(\lambda)$ is the generating function of the
integrals of motion: ${\mathfrak t}(\lambda)= \sum_n {t^{(n)} \over \lambda^n}$, where each $t^{(n)}$ is a constant of motion.

\subsubsection{The Toda Chain revisited}
\label{ssec:ALP_ExTodaDD}

\noindent 
We revisit the Toda chain based on the so-called dual description \cite{Sklyanin}. In this frame the Toda model is seen as one 
dimensional lattice model described by the semi-discrete auxiliary linear problem, with the corresponding Lax pair given by
\begin{equation}
	L_j(\lambda) = \lb \begin{matrix} \lambda - p_j & \e{q_j} \\ -\e{-q_j} & 0 
\end{matrix} \rb, ~~~~~~ A_j(\lambda) = \lb \begin{matrix} \lambda & \e{q_j} 
\\ -\e{-q_{j - 1}} & 0 \end{matrix} \rb.\label{eq:ALP_ExTodaDDLA}
\end{equation}
It is convenient for our purposes here to express the $L$-operator as $ L_j = \lambda D + \tilde{L}_j $, where:
\begin{equation}
	D = \lb \begin{matrix} 1 & 0 \\ 0 & 0 \end{matrix} \rb, ~~~~~~ \tilde{L}_j = \lb \begin{matrix} -p_j & \e{q_j} \\ -\e{-q_j} & 0 \end{matrix} \rb. \nonumber
\end{equation}
As explained in the preceding subsection to obtain the integrals of motion we need to expand the transfer matrix ${\mathfrak t}$ in powers of ${1\over \lambda}$. 
Let us first consider the expansion of the monoromy matrix:
\begin{align}
	T(\lambda) &= (\lambda  D_N + \tilde{L}_N) \ldots (\lambda D_1 + \tilde{L}_1) \nonumber \\
	&= \lambda^N \Big (  D_N \ldots D_1 + {1\over \lambda} \sum_{j = 1}^{N} D_N \ldots D_{j+1}\tilde{L}_j D_{j - 1} \ldots D_1 \nonumber\\
          &+ {1\over \lambda^{2}} \sum_{j >i } D_{N} \ldots  D_{j+1} \tilde{L}_j  D_{j-1} \ldots D_{i+1} \tilde L_{i} D_{i-1} \ldots D_1+ \ldots\Big ) \label{mono2}
\end{align}
and the transfer matrix reads 
\begin{equation}
{\mathfrak t}(\lambda) = tr \big (T(\lambda)\big ) = \lambda^N \Big (t^{(0)}+ {1\over \lambda} t^{(1)} + {1\over \lambda^2} t^{(2)} +  \ldots \Big ).  \label{transfer2}
\end{equation} 
From the last two expressions (\ref{mono2}), (\ref{transfer2}) and after some cumbersome, but straightforward calculations we read the first couple of integrals of motion
\begin{align}
& t^{(1)} =  - \sum_{j = 1}^{N} p_j  \nonumber\\
& t^{(2)} = \sum_{j>i=1}^N p_{j} p_i + \sum_{j=1}^N e^{q_{j+1} - q_j}.
\label{eq:ALP_ExTodaDDP}
\end{align}
Notice that $t^{(2)}$ is a {\it non-local} quantity, i.e. it involves interactions among all the sites of the one dimensional chain. 
In general, to obtain the {\it local} conserved quantities i.e. quantities that only involve interactions among neighbor sites one needs to expand:
$ ln \big( {\mathfrak t}(\lambda)\big) = \sum_{n} {I_n \over \lambda^n} $,
then one obtains for the first two local integral of motion,
\begin{eqnarray}
&& I_1 = t^{(1)} \nonumber\\
&& I_2= t^{(2)} - {(t^{(1)})^2 \over 2}.
\end{eqnarray}
The first quantity $I^{(1)}$  is simply the total momentum (\ref{eq:ALP_ExTodaDDP}), whereas the second charge is the local Toda Hamiltonian with periodic boundary conditions, 
(${\cal O}_{N+j} = {\cal O}_j$, where ${\cal O}_j \in \{q_j,\ p_j\}$):
\begin{equation}
	H = \frac{1}{2} \sum_{j = 1}^{N} p_j^2 + \sum_{j = 1}^{N} \e{q_j - q_{j - 1}}. \label{eq:ALP_ExTodaDDH}
\end{equation}
Note that the only difference compared to the expression we obtained via the {\it dual} description in the previous subsection is that now we consider periodic 
boundary conditions, whereas in the previous description we dealt with the open case (\ref{eq:ALP_ExTodaH}).

Using the semi-discrete zero curvature condition for the Lax pair (\ref{eq:ALP_dZCC}) we recover the familiar equations of motion for the Toda chain
\begin{eqnarray}
	&&\dot q_j= p_j   \nonumber \\
	&& \dot p_j = \e{q_{j+ 1} - q_j} - \e{q_j - q_{j - 1}}.\nonumber
\end{eqnarray}

\subsubsection{The Discrete Non-Linear Schr\"odinger model}
\label{ssec:ALP_ExDNLS}

\noindent 
We have looked a lot at the Toda chain in this section, so let us consider another prototypical integrable model, 
namely the discrete non-linear Schr\"odinger model (DNLS) \cite{KunduRagnisco, Ablo}. 
The Lax pair for this model is given by:
\begin{eqnarray}
	&&L_j(\lambda) = \lb \begin{matrix} \lambda +\mathbb{N}_j & x_j \\ X_j & 1 \end{matrix} \rb \label{eq:ALP_ExDNLSL}\nonumber \\
	&&A_j(\lambda) = \lb \begin{matrix} \lambda^2 - x_j X_{j - 1} & \lambda x_j - x_j \mathbb{N}_j + x_{j + 1} \\ \lambda X_{j - 1} -X_{j - 1} 
\mathbb{N}_{j - 1} +X_{j - 2} & x_j X_{j- 1} \end{matrix} \rb, \label{eq:ALP_ExDNLSA}
\end{eqnarray}
where we define  $ \mathbb{N}_j = 1 + x_j X_j $.  

As in the previous example, we are going to derive the conserved quantities of the system by expanding $ \ln{\mathfrak{t}} $ in powers of $ \lambda^{-1} $, 
except this time we shall find the first three. Splitting the $ L $ matrix up into $ L_j = \lambda D + \tilde{L}_j $ (where $ D $ is the same as it was in the previous 
example and $ \tilde{L}_j $ can be read off from the above definition of $ L_j $), $ T(\lambda) $ takes the same form as it did in that example, as does $ \ln{\mathfrak{t}} $. Therefore, the first integral of motion can be calculated as in the previous section. We keep here terms up to third order in the expansion, then after some lengthy 
calculations, in the spirit of the previous example we obtain the first few local integrals of motion of the DNLS model
\begin{eqnarray}
&&I_1= \sum_{j = 1}^{N} \mathbb{N}_j \label{eq:ALP_ExDNLSN} \nonumber\\
&& I_2= \sum_{j = 1}^{N} x_j X_{j - 1} - \frac{1}{2} \sum_{j = 1}^{N} \mathbb{N}_j^2 \nonumber\\
&& I_3= \sum_{j = 1}^{N} x_j X_{j - 2} - \sum_{j = 1}^{N} \lb \mathbb{N}_j + \mathbb{N}_{j - 1} \rb x_j X_{j - 1} + \frac{1}{3} \sum_{j = 1}^{N} \mathbb{N}_j^3.
\label{eq:ALP_ExDNLSP} 
\end{eqnarray}

Via the semi-discrete zero-curvature condition (\ref{eq:ALP_dZCC}) for the NLS Lax pair (\ref{eq:ALP_ExDNLSA}) we also obtain the 
equations of motion i.e. the ODEs that describe the time evolution 
of the fields $x_i,\ X_j$:
\begin{eqnarray}
	&&\dot x_j = x_j x_{j + 1} X_j - x_{j + 1} \lb \mathbb{N}_j + \mathbb{N}_{j + 1} \rb + 
x_j \mathbb{N}_j^2 + x_j^2 X_{j - 1} + x_{j + 2} \label{eq:ALP_ExDNLSxEvo} \nonumber\\
	&& \dot X_j = x_j X_{j - 1} X_j + X_{j - 1} \lb \mathbb{N}_j + \mathbb{N}_{j - 1} \rb-
 X_j \mathbb{N}_j^2 - x_{j + 1} X_j^2 - X_{j - 2}. \label{eq:ALP_ExDNLSXEvo}
\end{eqnarray}

\subsection{Continuous systems Lax pair}
\label{sec:CTS}

\noindent 
In the preceding subsection we discussed the class of one-dimensional semi-discrete integrable systems and we derived
their time evolution via the discrete zero curvature condition. In the present subsection we focus on
continuous space and time integrable systems. Again the key object in our construction is the continuous Lax pair  $\big(U,\ V \big )$, 
which satisfies the continuous auxiliary linear problem (see e.g. \cite{FT})
\begin{eqnarray}
&& \partial_x \Psi(x,t,\lambda)= U(x,t,\lambda) \Psi(x,t,\lambda)  \nonumber\\ 
&& \partial_t \Psi(x,t,\lambda)= V(x,t,\lambda) \Psi(x,t,\lambda). \label{eq:CTS_ALP}
\end{eqnarray}
$U,\ V$ can be in general $d \times d$ matrices depending on some fields and some spectral parameter.
Cross-differentiating the two equations above we obtain the continuous zero curvature condition
\begin{equation}
\partial_t U - \partial_xV+\big [ U,\ V \big ]=0. \label{emcont}
\end{equation}

The continuous analogue of the discrete monodromy matrix (\ref{mono1})  derived in the previous subsection is given as
\begin{equation}
T(x, y, t, \lambda) ={\cal P} \text{exp}\lb \int_{y}^{x}U(\xi) d \xi \rb, ~~~~~x\geq y \label{eq:CTS_T}
\end{equation}
and it is a solution of the $x$-part of the continuous auxiliary linear problem. The latter expression is a {\it path ordered} exponential, which is formally defined as
\begin{eqnarray}
&&{\cal P} \mbox{exp}\Big (  \int_{y}^{x} U(\xi, t) d \xi \ \Big )= \sum_{n=0}^{\infty} \int_{y}^xdx_n U(x_n)  
\int_y^{x_n }dx_{n-1}U(x_{n-1})   \ldots \int_y^{x_2} dx_{1} U(x_{1})\nonumber\\
&& x\geq x_n \geq x_{n-1} \ldots \geq x_1 \geq y.
\end{eqnarray}
The continuous monodromy matrix can be obtained as a suitable continuum limit of the discrete  one (\ref{mono1}) 
(we refer the interested reader to \cite{AvanDoikouSfetsos} for a detailed discussion and proof).

We consider that the system runs over some region $\big [-L,\  L \big ]$, and define the transfer matrix as in the discrete case,
$ \mathfrak{t}(t, \lambda) = tr\big ( T(L, -L, t, \lambda) \big) $. Given that this definition of $ \mathfrak{t}$ is the continuous limit of its discrete analogue, 
the continuous $ \mathfrak{t} $ will naturally be an integral of motion, 
and can be expanded in powers of $ \lambda^{-1}$ to provide an infinite tower of integrals of motion $ t^{(n)} $. 
To obtain the local integrals of motion, as in the discrete case, we need to consider the formal series expansion of  $ln \big ( {\mathfrak t}(\lambda) \big )$.

\subsubsection{Riccati equation $\&$ conserved quantities: AKNS system}

\noindent 
Our main aim now is the derivation of conserved quantities for continuous systems via the solution of an associated Riccati equation. To illustrate the computational details 
of this process we employ the prototypical AKNS system \cite{AKNS}.

We focus on the continuous monodromy matrix, and the fact that it is a solution of the $x$-part of the auxiliary linear problem (\ref{eq:CTS_ALP}).
We use the standard decomposition for the monodromy matrix (see also \cite{FT}):
\begin{equation}
	T(x, y, t, \lambda) = \big (\mathbb{I} + W(x, t, \lambda) \big) \e{Z(x, y, t, \lambda)} \big ( \mathbb{I} + W(y, t, \lambda) \big)^{-1}, \label{eq:CTS_TWZ}
\end{equation}
where $ W $ is a purely anti-diagonal matrix and $ Z $ is a purely diagonal one. Also, the matrix $ U $ is split into its diagonal and anti-diagonal parts, 
$ U = U_D + U_A $.

To determine the $ W $ and $ Z $ matrices, let us insert this expression back into Eq. \eqref{eq:CTS_ALP}:
\begin{align}
	&\pdiff{}{x} \Big ( \big( \mathbb{I} + W \big) \e{Z} \Big )= \big ( U_D + U_A \big) \big( \mathbb{I} + W \big ) \e{Z} \nonumber \\
	&\pdiff{W}{x} + \big( \mathbb{I} + W \big) \pdiff{Z}{x} = U_D + U_D W + U_A + U_A W. \nonumber
\end{align}
Splitting this into its diagonal and anti-diagonal components gives rise to two equations:
\begin{eqnarray}
	&&\pdiff{W}{x} + W \pdiff{Z}{x} = U_D W + U_A \nonumber\\
&&\pdiff{Z}{x} = U_D + U_A W,
 \label{eq:CTS_WZDiag} 
\end{eqnarray}
and inserting the second equation into the first  in place of the $ \pdiff{Z}{x}$ we obtain the fundamental Riccati equation for $W$
\begin{eqnarray}
	\pdiff{W}{x} + \big [ W,\ U_D \big ] + W U_A W - U_A = 0. \label{eq:CTS_WZAntiDiag}
\end{eqnarray}
Finally, as both $ W $ and $ Z $ are functions of the spectral parameter $ \lambda $ they can be expanded as (see also \cite{FT})
\begin{eqnarray}
&& W(x, t, \lambda) = \sum_{n = 1}^{\infty} { W^{(n)}(x, t) \over \lambda^{n} }, \nonumber\\
&&  Z(x, y, t, \lambda) = \sum_{n = -1}^{\infty}  {Z^{(n)}(x, y, t) \over \lambda^{n} }, \label{series}
 \label{eq:CTS_WZExpansion}
\end{eqnarray}
then  inserting these into our equations (\ref{eq:CTS_WZDiag}), (\ref{eq:CTS_WZAntiDiag}) allows the derivation of recursion relations 
for $ W^{(n)} $ and $ Z^{(n)}$ depending on the specific form of the $U$-operator.

\begin{itemize}

\item  {\bf The AKNS system}\\
We focus now on the AKNS system \cite{AKNS} for which the $U$-operator consists of the following diagonal and anti-diagonal parts
\begin{equation}
	U_D = {\lambda \over 2} \sigma, \qquad\qquad U_A = \lb \begin{matrix} 0 & 
\hat u\\  u& 0 \end{matrix} \rb, \label{eq:CTS_ExNLSU}
\end{equation}
where $\sigma = \mbox{diag}(1,\ -1)$.  It is worth noting that the AKNS system is a generic system that gives rise to 
many fundamental integrable models, more precisely:
\begin{enumerate}
\item {\tt Non-linear Schr\"odinger}:
$\hat u = u^*$
\item {\tt mKdV $\&$ sine-Gordon}:
$\hat u = u$
\item {\tt KdV}: $u=1$.
\end{enumerate}

Our main aim now is to identify the terms $W^{(n)},\ Z^{(n)}$ of the series expansion, and hence the conserved quantities. 
Recalling the generic equations for the diagonal and anti-diagonal parts 
(\ref{eq:CTS_WZDiag}), (\ref{eq:CTS_WZAntiDiag}), and taking into consideration the 
$\lambda$ series expansion (\ref{series}) we can derive the following recursion Riccati type relations for the anti-diagonal part:
\begin{eqnarray}
	&&W^{(1)} \sigma = U_A \nonumber \\
	 &&W^{(n + 1)}\sigma= -\pdiff{W^{(n)}}{x} - \sum_{m = 1}^{n - 1} W^{(n)} U_A W^{(n - m)}, \label{Ric1}
\end{eqnarray}
and for the diagonal part:
\begin{eqnarray}
	&&\pdiff{Z^{(-1)}}{x} = U_D \label{eq:CTS_ExNLSZm1} \\
	&&\pdiff{Z^{(n)}}{x} = U_A W^{(n)}. \label{eq:CTS_ExNLSZn}
\end{eqnarray}

It is then straightforward to evaluate the first few terms of the series expansion for the anti-diagonal part via the fundamental recursion relations above,
\begin{eqnarray}
&&W^{(1)} = \lb \begin{matrix} 0 & -\hat u \\u & 0 \end{matrix} \rb, 
~~~~W^{(2)} = \lb \begin{matrix} 0 & -\partial_x \hat u \\ -\partial_x u & 0 \end{matrix} \rb  \label{eq:CTS_ExNLSW1Mat}\\
	&&W^{(3)} = \lb \begin{matrix} 0 &-\partial_x^2 \hat u + u \hat u^2 \\ \partial_x^2  u -\hat  u u^2 & 0 \end{matrix} \rb \label{eq:CTS_ExNLSW3Mat}
\end{eqnarray}
and so on. Now that we have at our disposal the $ W^{(n)} $, we can use them to calculate the corresponding $ Z^{(n)} $ via Eqs. (\ref{eq:CTS_ExNLSZm1}), (\ref{eq:CTS_ExNLSZn}). 
Integrating over the interval $\big [-L,\  L\big ]$ these can be readily read off as
\begin{eqnarray}
	&&Z^{(-1)} = \lb \begin{matrix}L & 0 \\ 0 & -L \end{matrix} \rb \label{eq:CTS_ExNLSZm1Mat} \\
	&&Z^{(1)} = \lb \begin{matrix} \int_{-L}^{L} u \hat u\ dx  \\ 0 &- \int_{-L}^{L} u \hat udx \end{matrix} \rb \label{eq:CTS_ExNLSZ1Mat} \\
	&&Z^{(2)} = \lb \begin{matrix} \int_{-L}^{L} \hat u \partial_x u\ d x & 0 \\ 0 & -\int_{-L}^{L} u \partial_x \hat u \ dx \end{matrix} \rb 
\label{eq:CTS_ExNLSZ2Mat} \\
	&&Z^{(3)} = \lb \begin{matrix}\int_{-L}^{L} \big (\hat u \partial_x^2 u - (\hat u u )^2 \big )\ dx 
& 0 \\ 0 & \int_{-L}^{L} \big (- u \partial_x^2 \hat u + (\hat u u )^2 \big )\ dx \end{matrix} \rb. \label{eq:CTS_ExNLSZ3Mat}
\end{eqnarray}
With the knowledge of $ Z $ we will be able to derive the integrals of motion by expanding $ \ln{\mathfrak{t}} = \ln{\tr{\e{Z}}} $ in powers of $ \lambda $. 
The elements of the exponential of a diagonal matrix are just the exponentials of its elements, i.e. $ (\e{Z})_{ii} = \e{Z_{ii}} $, so this becomes:
\begin{equation}
	\ln{\mathfrak{t}} = \ln{\e{Z_{11}} + \e{Z_{22}}} \nonumber
\end{equation}

We consider one of two limits here, either $ \lambda \to \infty $ or $ \lambda \to -\infty $. 
In both limits, the leading order term in the exponentials will be 
$ \lambda Z_{ii}^{(-1)} $, but we have already calculated these in Eq. \eqref{eq:CTS_ExNLSZm1Mat}. 
Therefore, looking at the $ \lambda \to \infty $ limit, the leading order terms are
$ \e{\lambda L} $ and $ \e{-\lambda L} $ for $ \e{Z_{11}} $ and $ \e{Z_{22}} $ respectively.
Using the limit $ \lambda \to \infty $, $ \e{Z_{22}} \to 0 $, then the integrals of motion are just 
read off from the expansion of
\begin{equation}
	\ln{\mathfrak{t}} = \lambda Z_{11}^{(-1)} + \lambda^{-1} Z_{11}^{(1)} + \lambda^{-2} Z_{11}^{(2)} + ... \nonumber
\end{equation}
 which gives the first three as (ignoring the order $ \lambda $  term, as this is trivially an integral of motion):
\begin{eqnarray}
	&& I^{(1)} =\int_{-L}^{L} u \hat u\ dx  \nonumber\\
	&& I^{(2)} =  \int_{-L}^{L} \hat u \partial_x u\ d x\nonumber\\
	&& I^{(3)} =\int_{-L}^{L} \big (\hat u \partial_x^2 u - (\hat u u )^2 \big )\ dx, \label{integrals1}
\end{eqnarray}
and are identified with the number of particles, the momentum, and the Hamiltonian of the system respectively.

\item {\bf Equivalent description}

We describe in what follows an equivalent, albeit a bit more compact, way of deriving the conserved quantities, based 
also on the solution of the underlying Riccati equation emerging from the linear auxiliary problem.

The auxiliary function for our example here  is expressed as a column two-vector: $\Psi =  \lb \begin{matrix}
		\Psi_1 \\
		\Psi_2
	\end{matrix} \rb$.
Consider now the $x$-part of the auxiliary problem (\ref{eq:CTS_ALP}):
\begin{eqnarray}
&& \partial_x\Psi_1 = {\lambda \over 2}\Psi_1 +\hat u \Psi_2, \nonumber\\
&&   \partial_x\Psi_2 = -{\lambda \over 2}\Psi_2 + u \Psi_1,
\end{eqnarray}
and define the element $\Gamma = \Psi_2\ \Psi_1^{-1}$. Then from the latter expressions 
one arrives at the Riccati equation obeyed by $\Gamma$:
\begin{equation}
\partial_x \Gamma = u -\lambda\Gamma -\Gamma\ \hat u\  \Gamma. \label{Ric}
\end{equation}
It is worth noting, in comparison with the previous description based on the decomposition of the monodromy matrix, that  $\Gamma$ is essentially 
the analogue of the element $W_{21}$ and $\Psi_1$ is the analogue of $e^{Z_{11}}$, a fact that underlines the equivalence of the two descriptions.
Similarly, we could have defined $\hat \Gamma = \Psi_1 \Psi_2^{-1}$, then $\hat \Gamma$ would be the equivalent of $W_{12}$ and $\Psi_2$ 
the equivalent of  $e^{Z_{22}}$.

The next important task is to identify the element $\Gamma$.
This can be achieved by expressing $\Gamma$ in a formal power series expansion $\Gamma = \sum_{k} {\Gamma^{(k)} \over \lambda^k}$ 
and solving the Riccati equation at each order.
Then (\ref{Ric}) reduces to $\Gamma^{(1)} = u$, and
\begin{eqnarray}
\partial_x \Gamma^{(k)}= - \Gamma^{(k+1)} - \sum_{l=1}^{k-1}\Gamma^{(l)}\  \hat u\  \Gamma^{(k-l)}, ~~~~k >0.
\end{eqnarray}
The latter expression is the equivalent of (\ref{Ric1}). Let us report below the first few terms of the expansion
\begin{eqnarray}
&& \Gamma^{(1)}= u , ~~~~\Gamma^{(2)}=-\partial_x u, ~~~~ \Gamma^{(3)}=\partial_x^2u -u\hat u u,~~\ldots
\end{eqnarray}

The next natural step is to identify the associated conserved quantities by means of the auxiliary linear problem relations.
Let us express the time components of the Lax pair $\big ( U,\ V^{(n)} \big )$ as $V^{(n)} =   \lb \begin{matrix}
		\alpha_n&  \beta_n\\
		\gamma_n &\delta_n
	\end{matrix} \rb$, (associated to the time $t_n$) and also recall for both the $x$ and $t$-parts of the linear problem:
\begin{eqnarray}
&& \partial_x \Psi_1\ \Psi_1^{-1} =  {\lambda \over 2} + \hat u \Gamma,   \nonumber\\
&&  \partial_{t_n} \Psi_1\ \Psi_1^{-1} = \alpha_n +\beta_n \Gamma.
\end{eqnarray}
After cross differentiating the equations above we conclude
\begin{equation}
\partial_{t_n}\big (\hat u \Gamma\big ) = \partial_x\big (\alpha_n + \beta_n \Gamma\big ). \label{basis}
\end{equation}
From the latter expression it is clear that the quantities 
\begin{equation}
I^{(k)} = \int_{\mathbb R} dx\ \hat u(x) \Gamma^{(k)}(x), \label{block}
\end{equation}
are automatically conserved (see also \cite{FT}), where we have assumed vanishing boundary conditions at $\pm \infty$. Substituting the computed values of 
$\Gamma^{(k)}$ for $ k= 1,\ 2, \ 3, \ldots$ we recover the integrals of motion (\ref{integrals1}), as expected  (consider $L \to \infty$).

\end{itemize}

\section{Darboux-B\"acklund  Transformations}
\label{sec:Dbx}
\noindent
After having discussed in detail the systematic construction of conserved quantities via the Lax pair formulation for both discrete
and continuous integrable systems we come to the other main objective of this presentation, which is the solution of integrable ODEs/PDEs 
via general algebraic schemes known as the Darboux-B\"acklund transformation (BT) or the generalized Zakharov-Shabat dressing scheme involving differential 
and integral operators as dressing transformations.

\subsection{Discrete Darboux-BT}

\noindent 
Let us first focus on the issue of the discrete Darboux-B\"acklund transformation (BT) in the context of semi-discrete integrable models. 
Recall the auxiliary linear problem for such systems
\begin{eqnarray}
&& \Psi_{j + 1}(t, \lambda)= L_j (t, \lambda)\ \Psi_j (t, \lambda), \nonumber\\
&& \partial_t \Psi_j (t, \lambda)= A_j(t, \lambda)\ \Psi_j(t, \lambda). \label{eq:Dbx_DALP}
\end{eqnarray}
We introduce the so-called  Darboux matrix \cite{Darboux}, a \nbyn{d} matrix $ {\mathrm G} $ that transforms the auxiliary 
function $\Psi_j$: $\tilde{\Psi}_j = {\mathrm G}_j \Psi_j $  and leaves the form of the integrals of motion invariant. 
The main assumption now is that $ \tilde{\Psi}_j,\ \tilde L_j,\ \tilde A_j$ also satisfy the auxiliary linear problem,
\noindent where $ \tilde{L}_j $ and $ \tilde{A}_j $ comprise the Lax pair associated to this new auxiliary linear problem. 
Inserting the definition of $ \tilde{\Psi}_j $ 
into the first of these relations:
\begin{align}
	&  {\mathrm G}_{j + 1}\ \Psi_{j+ 1}  = \tilde{L}_j\  {\mathrm G}_J \Psi_j  \nonumber \\
	& {\mathrm G}_{j+ 1} \lb L_j \Psi_j \rb = \tilde{L}_j\ {\mathrm G}_j \Psi_j, \nonumber
\end{align}
and inserting the definition into the second relation we conclude
\begin{align}
	&\partial_t  \lb {\mathrm G}_j \Psi_j \rb = \tilde{A}_j\ \lb {\mathrm G}_j \Psi_j \rb \nonumber \\
	& \partial_t {\mathrm G}_j\ \Psi_j + {\mathrm G}_j\ \lb A_j \Psi_j \rb = \tilde{A}_j\ \lb {\mathrm G}_j \Psi_j\rb. \nonumber
\end{align}
These then give two expressions relating the original Lax pair $ (L_j,\  A_j) $ with the new Lax pair $ (\tilde{L}_j,\  \tilde{A}_j) $:
\begin{eqnarray}
	&& {\mathrm G}_{j + 1} L_j = \tilde{L}_j {\mathrm G}_j \label{eq:Dbx_DTFEqL} \nonumber\\
	&& \partial_t {\mathrm G}_j = \tilde{A}_j {\mathrm G}_j - {\mathrm G}_j A_j. \label{eq:Dbx_DTFEqA}
\end{eqnarray}

Given a Darboux matrix that satisfies the above construction, and solving these two equations we can actually obtain the B\"acklund 
transformation that connects solutions between the two systems described by $ (L_j,\  A_j) $ and $ (\tilde{L}_j,\  \tilde{A}_j) $. 
These can be different solutions of the same PDE (auto-BT) or solutions of different PDEs (hetero-BT), depending on the forms of the two Lax pairs.

\subsubsection{The Toda Chain}
\label{ssec:Dbx_ExToda}

\noindent 
To illustrate  how the BT works we  now focus on our typical discrete example of the Toda chain, 
and derive solutions of the equations of motion via the BT relations. We recall the Lax pair for the model
\begin{equation}
	L_j = \lb \begin{matrix} \lambda - p_j & \e{q_j} \\ -\e{-q_j} & 0 \end{matrix} \rb, ~~~~~~~ A_j = 
\lb \begin{matrix} \lambda & \e{q_j} \\ -\e{-q_{j - 1}} & 0 \end{matrix} \rb \label{eq:Dbx_ExTodaLA}
\end{equation}
\noindent and we chose to consider the following Darboux matrix
\begin{equation}
	{\mathrm G}_j = \lb \begin{matrix} \lambda + {\mathbb N}_j & X_j \\ Z_j & 1 \end{matrix} \rb, \label{eq:Dbx_ExTodaM}
\end{equation}
where  ${\mathbb N}_j = \Theta +X_j Z_j $.

The discrete spatial part of the BT relations produces
\begin{eqnarray}
&& X_j= e^{q_j}, ~~~Z_{j+1} = - e^{-\tilde q_j} \nonumber\\
&& {\mathbb N}_{j+1} q^{ q_j}= -\tilde p_j X_j + e^{\tilde q_j}, ~~~~ e^{-\tilde q_j}{\mathbb N}_j = e^{-q_j} +Z_{j+1} p_j,
\end{eqnarray}
plus two  extra compatibility conditions emerging from the diagonal entries, which are omitted for brevity. The $t$-part of the BT relations provides
\begin{eqnarray}
\dot X_j = e^{\tilde q_j} - {\mathbb N}_j e^{q_j}, ~~~~\dot Z_j = -e^{-\tilde q_{j-1}} {\mathbb N}_j + e^{-q_{j-1}}.
\end{eqnarray}

Gathering the information provided by the $x$ and $t$ parts of the BT above we conclude
\begin{eqnarray}
&& \dot q_j= p_j = e^{\tilde q_j - q_j} +  e^{q_j - \tilde q_{j-1}} -\Theta \nonumber\\
&& \dot {\tilde q}_j= \tilde p_j = e^{\tilde q_j - q_j} +e^{q_{j+1} - \tilde q_j} - \Theta,  \label{basic3}
\end{eqnarray}
and consequently  one can show that $q_j,\ \tilde q_j$ satisfy the Toda equations of motion. 

We focus our attention on deriving solutions of the Toda equations. To achieve this we consider a trivial solution of the equations,
i.e. let $\tilde p_j = \tilde q_j =0$ (or equivalently one may consider $q_j = p_j =0$), then equations (\ref{basic3}) become
\begin{eqnarray}
&& 0= e^{-q_j} + e^{q_{j+1}} -\Theta \label{A} \\
&&  p_j = e^{-q_j} + e^{q_j} -\Theta. \label{B}
\end{eqnarray}
We can now readily obtain the stationary solution i.e. consider only the time independent  equation (\ref{A}). 
It is convenient to parametrize as follows (see also \cite{Sklyanin} and references therein),
\begin{equation}
\Theta = 2 \cosh \eta, ~~~~e^{q_0}= {\cosh{\xi + \eta} \over \cosh{\xi}},
\end{equation}
then the solution of the difference equation (\ref{A}) is given as
\begin{equation}
e^{q_j} = {\cosh{\xi + \eta (j+1)} \over \cosh{\xi + \eta j}}.
\end{equation}

The second equation (\ref{B}) yields the time dependence of the solution. Compare the solution above  
with the traveling wave solution introduced earlier in the text.

\subsection{Continuous Darboux-BT}
\label{ssec:CtsDbx}

\noindent 
In the continuous case, the process is much the same, except that we start from the continuous auxiliary linear problem. 
We introduce a Darboux matrix $ {\mathrm G} $ analogously to the discrete case, i.e. $\tilde \Psi(x) = {\mathrm G}(x)  \Psi(x)$, where $\tilde \Psi$ 
is the transformed auxiliary function. As in the semi-discrete case, this allows us to find a pair of continuous relations between the Darboux matrix and the original and 
``transformed'' Lax pairs, $ (U, V) $ and $ (\tilde{U}, \tilde{V}) $. 

Indeed, after differentiating the transformed auxiliary problem we obtain the fundamental BT 
relations for the continuous integrable system:
\begin{eqnarray}
&& \partial_x{\mathrm G}(\lambda, x,t ) = \tilde{U}(\lambda, x,t ) {\mathrm G}(\lambda, x,t )  - 
{\mathrm G}(\lambda, x,t ) U(\lambda, x,t ) \label{eq:Dbx_CtsTFEqU} \nonumber\\
&& \partial_t {\mathrm G}(\lambda, x,t ) = \tilde{V}(\lambda, x,t ){\mathrm G}(\lambda, x,t ) - 
{\mathrm G}(\lambda, x,t ) V(\lambda, x,t ). \label{eq:Dbx_CtsTFEqV}
\end{eqnarray}
Solving the pair of relations above allows the derivation of the Darboux matrix and also provides relations 
between the two different solutions of the associated integrable PDEs. 
This will become transparent in the various  examples discussed below.

\subsubsection{The sinh-Gordon Model: auto-BT}
\label{ssec:Dbx_ExsG}

\noindent 
A significant physical example is considered in this subsection, i.e. the sinh-Gordon model. 
This is a relativistic model as opposed to the NLS model, which is associated to a typical reaction-diffusion Hamiltonian. 
The sinh-Gordon Lax pair reads as (see e.g. \cite{FT}):
\begin{eqnarray}
	&&U(\lambda, x, t) = \frac{1}{2} \lb \begin{matrix} -\partial_t w & \sinh{\lambda + w} \\ 
\sinh{\lambda - w} & \partial_t w \end{matrix} \rb, \nonumber\\ 
& &V(\lambda, x,t) =
 \frac{1}{2} \lb \begin{matrix} -\partial_x w & \cosh{\lambda + w} \\ \cosh{\lambda - w} & \partial_x w
\end{matrix} \rb \label{eq:Dbx_ExsGUV}
\end{eqnarray}
where we define $w= {\beta \over 2} \phi$, $\beta$ is the coupling constant of the model, and 
the sinh-Gordon field  $\phi$ naturally depends on $x,\ t$.
The zero curvature condition for the Lax pair above 
provides the sinh-Gordon equation
\begin{equation}
\partial_t^2 \phi - \partial_x^2 \phi = \beta^{-1}\sinh{\beta \phi}. \label{SG}
\end{equation}
Note that the sine-Gordon model emerges by simply setting $\phi \to {\mathrm  i} \phi$.

We will be using the type \rm{I} Darboux matrix, given as  (there is also a type \rm{II} matrix, 
which can be written as the product of two type \rm{I} matrices)
\begin{equation}
	{\mathrm G} = \lb \begin{matrix} Z & \e{\lambda-\Theta} X^{-1} \\ \e{\lambda-\Theta} X & Z^{-1} \end{matrix} \rb. \label{eq:Dbx_ExsGM}
\end{equation}
Inserting this into the space equation\footnote{Due to the fact that the sinh-Gordon is a relativistic model it suffices to only consider the space 
(or time) part of (\ref{eq:Dbx_CtsTFEqV}) in order to obtain the BT.} 
\eqref{eq:Dbx_CtsTFEqU} we can immediately read off expressions for $ X,\ Z$:
\begin{equation}
	X = \e{-\frac{1}{2} \lb w + \tilde{w} \rb}, ~~~~ Z = \e{\frac{1}{2} \lb w - \tilde{w} \rb} \label{eq:Dbx_ExsGX}
\end{equation}
and we also obtain four more equations
\begin{align}
	&2 \partial_x Z =Z\big ( \partial_t w -  \partial_t \tilde w\big ) + \frac{e^{-\Theta}}{2} \big ( X^{-1} \e{w} -  \e{-\tilde{w}} X\big)  \nonumber \\
	& 2 \partial_x X  = \frac{e^{\Theta}}{2}\lb Z \e{-\tilde{w}} - Z^{-1}\e{-w} \rb + X  \partial_t \big (w+ \tilde w  \big ) \nonumber \\
	& 2 \partial_x X^{-1} = \frac{e^{\Theta}}{2} \lb Z^{-1}\e{\tilde{w}} - Z \e{w} \rb - X^{-1} \partial_t\big ( w - \tilde w \big ) \nonumber \\
	&2 \partial_x Z^{-1} =  Z^{-1}  \partial_t  \big (w -\tilde w \big ) \frac{e^{-\Theta}}{2} 
\e{\tilde{w}} X^{-1} + \frac{1}{4} X \e{-w}. \label{basic2}
\end{align}

Combining now the fundamental sets of relations above we conclude
\begin{align}
& \partial_x\lb w + \tilde w \rb = -\partial_t \lb w +\tilde w  \rb -\alpha\ \sinh{w - \tilde{w}}  \nonumber\\
& \partial_x \lb w - \tilde w \rb = \partial_t \big (w- \tilde w \big )+ \alpha^{-1}\sinh{w + \tilde{w}}, \label{eq:Dbx_ExsGBT2}
\end{align}
where $\alpha = e^{\Theta}$.  The equations above provide precisely the B\"acklund transformation of the sinh-Gordon model.
It is convenient to express the BT in light cone coordinates: $z = x+t,\ \bar z = x-t$,
\begin{align}
& \partial_z  \big ( w +\tilde w \big ) = \alpha\  \sinh{ \tilde w - w} \nonumber\\
& \partial_{\bar z} \big (w -\tilde w\big ) = \alpha^{-1} \sinh {w + \tilde w}. \label{BTS}
\end{align}
It is then straightforward by cross differentiating the above equations to show that both fields $w,\ \tilde w$ satisfy 
the sinh-Gordon equation (\ref{SG}). Moreover, the one soliton solution for the sinh-Gordon equation can be immediately identified 
when setting $ \tilde w =0$ (or $w =0$). Let $\tilde w =0$, then equations (\ref{BTS}) become simple to solve, and the solution is readily  
extracted 
\begin{equation}
\phi(z, \bar z )={ 2\over \beta} \mbox{ln}\Big ({1+Ae^{-\alpha z + \alpha^{-1} \bar z} \over 1- Ae^{-\alpha z + \alpha^{-1} \bar z}} \Big ),
\end{equation}
where $A$ is an integration constant.

One may construct multi-soliton solutions by either considering more general forms of BTs or by constructing
the so-called soliton lattice using {\it Bianchi's permutability theorem} \cite{Bianchi} (see also \cite{Drazin} and references therein).
The fundamental requirement in this setting is the commutation of fundamental BTs for different constants. Schematically, this idea can be demonstrated as follows
\begin{equation}
\text{\Large $ \begin{tikzcd}[row sep = small, column sep = large, ampersand replacement = \&]
	\- \& w_1^{\phantom{1}} \arrow[dr, "\lambda_2", end anchor = {north west}] \& \- \\
	w_0 \arrow[ur, "\lambda_1", start anchor = {north east}] \arrow[dr, swap, "\lambda_2", start anchor = {south east}] \& \- \& w_{12} = w_{21} \\
	\- \& w_2^{\phantom{2}} \arrow[ur, swap, "\lambda_1", end anchor = {south west}] \& \-
\end{tikzcd} $} \label{lattice}
\end{equation}
requiring  the commutativiy of the successive BTs  i.e. $w_{12} = w_{21} $. Let us more systematically see where this leads.
We recall the BT for the sinh-Gordon model, and focus on the $z$-part of the transform (equivalent findings emerge 
by considering the $\bar z$ part of the transform).
From (\ref{BTS}) we have
\begin{eqnarray}
&& \partial_z \big (w_i +w_0 \big ) = \alpha_i\ \sinh{w_i -w_0} \nonumber\\
&&   \partial_z \big (w_{ij} +w_{i} \big ) = \alpha_j\ \sinh{w_{ij} -w_i}, ~~~~i  \neq  j \in \{1,\ 2\}.
\end{eqnarray}
The aim now is to obtain the 2-soliton solution via the one-soliton solutions $w_1,\ w_2$. 
First we consider the trivial solution $w_0 =0$ 
as a starting solution, and also set $w_{12} = w_{21} = w$. Then from the equations above it follows
\begin{equation}
\alpha_1 \sinh {-w+w_2} - \alpha_2\ \sinh{-w+w_1} = \alpha_1 \sinh{w_1} - \alpha_2\ \sinh{w_2}.
\end{equation}
Solving the equation above for $w$ we obtain:
\begin{equation}
\Big ( e^{-w} + e^{-(w_1 +w_2)}\Big ) \Big ( \big (\alpha_1 e^{w_2} - \alpha_2 e^{w_1} \big) - e^{w}e^{w_1 +w_2}
\big (\alpha_1 e^{-w_2} - \alpha_2 e^{-w_1} \big ) \Big ) =0,
\end{equation}
and consequently
\begin{equation}
w =\mbox{ln} \Big ({ \alpha_1\ e^{{\Delta \over 2}} -\alpha_2\ e^{-{\Delta \over 2}} \over \alpha_1\ e^{-{\Delta \over 2}} -\alpha_2\ e^{{\Delta \over 2}}}\Big ),
\end{equation}
where $\Delta = w_2-w_1$. The latter expression provides a ``non-linear'' super-position for the solutions of the integrable PDE. 
We shall present a similar construction for the KdV equation in a subsequent section.

\subsubsection{The Liouville Model: hetero-BT}
\noindent 
We have seen so far examples of BTs that connect different solutions of the same integrable PDEs. 
However, one can consider BTs  that connect solutions of different integrable PDEs (hetero-BTs).
We  focus on a particular example of hetero-BT, and we find solutions of the Liouville equations via solutions of the massless free particle equation. 
Indeed, consider the following Lax pair associated to the Liouville equation
\begin{equation}
	\tilde U(\lambda) = \frac{1}{2} \left( \begin{matrix}
		-\tilde \pi & -2c \e{-\lambda + \tilde \phi} \\
		-2c\, \e{\lambda +\tilde \phi} &\tilde \pi
	\end{matrix} \right), ~~~~~~~
	\tilde V(\lambda) = \frac{1}{2} \left( \begin{matrix}
		-\partial_x \tilde \phi & 2c\e{-\lambda + \tilde \phi} \\
		2c\, \e{\lambda + \tilde \phi} &\partial_x \tilde \phi
	\end{matrix} \right).
\label{eq:LaxLiouv2}
\end{equation}
\noindent The equations of motion emerging from the zero curvature condition read as ($\tilde \pi = \partial_t \tilde \phi$)
\begin{equation}
	\partial_x^2 \tilde \phi - \partial_t^2 \tilde \phi- 4c^2 \e{2\tilde \phi} = 0. \label{eq:em1}
\end{equation}

\noindent The Lax pair for the free massless  theory is very simple and is given as
\begin{equation}
	U(\lambda) = -\frac{1}{2} \pi\ {\mathbb I}, ~~~~~~~
	V(\lambda) = -\frac{1}{2} \partial_x\phi\ {\mathbb I},
\label{eq:Laxfree}
\end{equation}
\noindent where  ${\mathbb I}$ is the $ 2 \times 2 $ identity matrix and the corresponding equation of motion is ($\pi = \partial_t \phi$)
\begin{equation}
	\partial_x^2 \phi - \partial_t^2 \phi =0. \label{eq:em2}
\end{equation}
\noindent It is clear that the equation of motion remains invariant if we multiply $U$ and $V$ with the same constant matrix.

We choose to consider the following Darboux-matrix:
\begin{equation}
	{\mathrm G}(\lambda,  \Theta )=  \left( \begin{matrix}
		A & X \e{-\lambda-\Theta}\\
		Z \e{\lambda+\Theta} & B
	\end{matrix} \right),
\end{equation}
\noindent where $\Theta$ is an extra free parameter (B\"acklund transformation parameter) and the elements $A,\  B,\  X,\  Z$ are to 
be determined via (\ref{eq:Dbx_CtsTFEqV}). Indeed, setting:
\begin{equation}
	A = X = \e{\frac{1}{2} (\tilde \phi -\phi)}, ~~~~Z = B = \e{\frac{1}{2} (\tilde \phi +\phi)},
\end{equation}
\noindent and using light cone coordinates $z = x + t,\ \bar{z} = x - t$ for convenience, we solve (\ref{eq:Dbx_CtsTFEqV}) and  obtain:
\begin{equation}
\begin{gathered}
	\partial_z(\tilde \phi -\phi) = -2c \e{\Theta} \e{ (\tilde \phi+\phi)}, \\
	\partial_{\bar z}(\tilde \phi +\phi) =-2c \e{-\Theta} \e{ (\tilde \phi-\phi)},
\end{gathered} \label{eq:lbt}
\end{equation}
\noindent which is the celebrated hetero-B\"acklund transformation for the Liouville theory.
It is easy to check via \eqref{eq:lbt} that the fields $\tilde \phi$ and $\phi$ satisfy the correct equations of motion \eqref{eq:em1} and \eqref{eq:em2} respectively. 

One can express solutions of the Liouville theory  in terms of solutions of the linear problem associated to the 
massless free relativistic particle. Indeed, due to the form of the free particle equation the solutions can be expressed as
$\phi(z, \bar z) = f(z) +\bar f(\bar z)$, then the BT relations become (we set for simplicity $\Theta =0$)
\begin{eqnarray}
&& e^{-(\tilde \phi + \bar f - f)}\partial_z(\tilde \phi -f) = -2c \e{ 2 f}, \nonumber\\
&& e^{-(\tilde \phi + \bar f - f)}\partial_{\bar z}(\tilde \phi +\bar f) =-2c  \e{ -2 \bar f },
\label{eq:lbt2}
\end{eqnarray}
which lead to:
\begin{equation}
e^{-(\tilde \phi+ \bar f - f) } = 2c \int^z e^{2f(\xi)}d\xi +2c \int^{\bar z} e^{2\bar f(\xi)}d\xi.
\end{equation}
Let $F(z) =  \int^z e^{2f(\xi)}d\xi $ and $\bar F^{-1}(\bar z)= -\int^{\bar z} e^{-2\bar f(\xi)}d\xi$, then the solution of the 
Liouville equation can be expressed in terms of $F,\ \bar F$ as
\begin{equation}
\tilde \phi(z,\bar z) = {1\over 2}\mbox{ln} \Big ( {\partial_z F\ \partial_{\bar z}\bar F \over (1-F \bar F)^2}\Big ).
\end{equation}
This concludes our brief discussion on  BTs associated to relativistic integrable models. Next we examine
the BT and dressing for the AKNS system.

\subsubsection{The AKNS system: BT $\&$ dressing}
\noindent 
We focus in this subsection on the AKNS (NLS-type) system (see e.g. \cite{Manakov}--\cite{DoFiSk}, \cite{Ablo} for the NLS model and generalizations),
we derive the BT relations and also perform the dressing process in order to derive the members of the AKNS hierarchy.
Recall that the BT transformation connects two solutions of the same (or distinct) PDEs and is identified provided the existence of given Lax pairs. 
In the dressing process, on the other hand the form of the so-called ``bare'' (free of fields) operators are given, and then via the Darboux-dressing process 
the Lax pairs for the associated  hierarchy are identified. 

Let us outline what we are going to achieve in this subsection.
We first employ a given  Lax pair $(U,\ V)$, e.g. the NLS one (\ref{eq:NLS_Lax}), and the Darboux matrix to identify 
the corresponding  B\"acklund transformation.  We then restrict our attention to the case of trivial solutions $u,\  \hat u =0$, identify the one-soliton solution, 
and we also perform the ``dressing''  process.  From this viewpoint the only input is the $U$-operator, together with the form of the bare operators, then the whole hierarchy of the 
$V$-operators can be identified via the dressing process. Note that in all the expressions below $x$ and $t$ dependence is implied, even when it is not explicitly stated.

For the NLS model the Lax pair  is given as
\begin{equation}
	U(\lambda, x,t) = \lb \begin{matrix}
		\frac{ \lambda}{2}& \hat   u \\
		 u & -\frac{ \lambda}{2}
	\end{matrix} \rb, ~~~~V(\lambda, x,t )=   \lb \begin{matrix}
		{\lambda^2 \over 2}-\hat u u & \lambda \hat u+\partial_x\hat u\\
		\lambda u-\partial_x u & - {\lambda^2 \over 2} + u \hat u
	\end{matrix}\rb,\label{eq:NLS_Lax}
\end{equation}
where the fields $\hat u,\  u$ depend on $x,\ t$. 
The equations of motion from the zero curvature condition lead to the NLS-type equation:
\begin{equation}
\partial_t u+\partial_x^2 u -2\hat u u^2=0.
\end{equation}
Similarly, for $\hat u$  ($t \to -t$).
The Darboux matrix is chosen to be of the form
\begin{equation}
{\mathrm G}(\lambda, x, t)=  \lb \begin{matrix}
		\lambda + {\cal A}(x, t) & {\cal B}(x, t) \\
		{\cal  C}(x, t) & \lambda +{\cal D}( x, t)
	\end{matrix} \rb = \lambda {\mathbb I} + {\cal K}. \label{GG}
\end{equation} 
This is the fundamental Darboux matrix, we shall see later in the text that the generic Darboux is expressed in a $\lambda$-power series. 
The BT connects two  different solutions of the underlying integrable PDEs, i.e the pair $U,\  V$ are associated to the fields $u,\ \hat u$, 
whereas $U_0,\  V_0$ are associated to the fields $u_0,\  \hat u_0$. 

The Darboux transform ${\mathrm G}$ is now applied on the  auxiliary function $\Psi_0$
\begin{equation}
\Psi(\lambda, x, t)= {\mathrm G}(\lambda, x,t )\  \Psi_0(\lambda, x, t), \label{Darboux2}
\end{equation}
yielding  (see also (\ref{generalDarboux})) the two primary Darboux-BT equations 
\begin{equation}
\partial_x{\mathrm G} = U\ {\mathrm G}-  {\mathrm G}\ U_0,
~~~~\partial_{t}{\mathrm G} = V\ {\mathrm G} -{\mathrm  G}\  V_0\ . \label{fundam2}
\end{equation}
The $x$-part  of (\ref{fundam2}) leads to the Darboux-BT relations ${(\cal D} = -{\cal A}$):
\begin{eqnarray}
&& {\cal B}= -(\hat u - \hat u_0), ~~~~~{\cal C}= u - u_0, \nonumber\\
&& \partial_x{\cal A} =\hat u {\cal C} - {\cal B}u_0 , ~~~~\partial_x {\cal D} =   u {\cal B}- {\cal C}\hat u_0, \nonumber\\
&& \partial_x  {\cal B} = \hat u{\cal  D}- {\cal A} \hat u_0 , ~~~~ \partial_x {\cal C} = u {\cal A}-  {\cal D} u_0. \label{basicxx}
\end{eqnarray}
For the dressing process  described next the trivial solution $u_0 = \hat u_0 =0$ is employed.
The $t$-part of the BT also gives the following:
\begin{eqnarray}
&& \partial_t {\cal B} =- \partial_x \big ( \hat u_0 + \hat u \big ) {\cal A} -\big (u \hat u + u_0 \hat u_0) {\cal B}\nonumber\\
&&  \partial_t {\cal C} =-\partial_x \big( u + u_0\big ) {\cal A} + \big (u \hat u + u_0 \hat u_0) {\cal C}.
\end{eqnarray}
We also require that the $\mbox{det} {\cal K}$ is a constant, which yields
\begin{equation}
{\cal A} = \sqrt{k^2+(u - u_0)(\hat u - \hat u_0)}. \label{basic2xx}
\end{equation}

Gathering all the relations above we end up with the BT for the NLS-type system
\begin{eqnarray}
&& \partial_x\big ( u - u_0 \big ) =  \big ( u + u_0 \big ){\cal A}\nonumber\\
&& \partial_t \big ( u -u_0 \big ) = -\partial_x\big ( u + u_0 \big ){\cal A} + {\cal X}\big ( u - u_0\big),
\end{eqnarray}
where we define
\begin{equation}
{\cal X} =  {1\over 2} \Big ( (u-u_0)(\hat u - \hat u_0) +  (u+u_0)(\hat u +\hat u_0) \Big)
\end{equation}
and  similarly for $\hat u,\ \hat u_0$
\begin{eqnarray}
&& \partial_x\big ( \hat u -\hat  u_0 \big ) =  \big ( \hat u +\hat  u_0 \big ){\cal A}\nonumber\\
&& \partial_t \big (\hat  u -\hat  u_0 \big ) = \partial_x\big ( \hat  u +\hat   u_0 \big ){\cal A} - 
{\cal X}\big ( \hat  u -\hat  u_0\big).
\end{eqnarray}
From the equations above one may show that both sets $u,\ \hat u$ and
$u_0,\ \hat u_0$ satisfy the NLS-type equations (we leave the explicit proof  as an exercise to the interested reader). 
Moreover, by considering  the trivial solutions $u_0 = \hat u_0 =0$ we can identify 
the soliton-type solution. Indeed, for $u_0 = \hat u_0 =0$ recall the fundamental relations for 
${\cal A}$ (\ref{basicxx}), (\ref{basic2xx}), which lead to a simple differential equation
\begin{equation}
\partial_x {\cal A} = -k^2 +{\cal A}^2\ \Rightarrow\ {\cal A}= {k (1 +e^{2kx + f(t)}) \over 1 - e^{2kx+f(t)}}.
\end{equation}
Then from these equations one can easily derive the solution for the fields $u,\ \hat u$ (this too is left as an exercise for the curious reader).

We keep focusing on the case where $u_0 = \hat u_0 =0$ and via the dressing process we are going to identify the 
``dressed'' quantities  $V^{(n)}$ of the hierarchy, assuming knowledge of the form of the $U$-operator as well as the bare operators. 
Let $U_0,\ V_0^{(n)}$ be the ``bare'' Lax pairs:
\begin{equation}
U_0(\lambda) = {\lambda \over 2}\sigma,~~~~V_0^{(n)}(\lambda)= {\lambda^{n} \over 2}\sigma,
\end{equation}
$\sigma = \mbox{diag}(1,\ -1)$. Also, the ``dressed'' time components of the Lax pairs can be expressed as formal series expansions
\begin{equation}
V^{(n)}(\lambda, x, t) = {\lambda^n \over 2}\sigma + \sum_{k=0}^{n-1}\lambda^k w^{(n)}_k(x, t),
\end{equation}
where the quantities $ w^{(n)}_k$ will be identified via the dressing transform.

By solving the $x$-part of equations (\ref{fundam2}) we obtain
\begin{equation}
\hat u =-{\cal  B}, ~~~u ={\cal C} ~~~\mbox{and} ~~~\partial_{x} {\cal K} =   \lb \begin{matrix}
		0 & \hat u\\
		u &0
	\end{matrix} \rb {\cal K}. \label{constr2b}
\end{equation}

From the time part of (\ref{fundam2}) we obtain a set of recursion relations, in exact 
analogy to the case of dressing via a differential operator, i.e.
\begin{eqnarray}
&& w^{(n)}_{n-1}= {1\over 2}\big [{\cal K},\ \sigma\big ], \nonumber \\
&& w^{(n)}_{k-1} =-w^{(n)}_{k}\  {\cal K}, ~~~~k\in\big \{1, 2, \ldots, n-1\big \}\nonumber\\
&& \partial_{t_n} {\cal K} = w^{(n)}_0\  {\cal K}. \label{recursion1}
\end{eqnarray}
We solve the latter recursion relations, and identify the first few time components of the Lax pairs:
\begin{eqnarray}
&& V^{(0)} = {1\over 2} \sigma, \nonumber\\
&& V^{(1)} = \lambda V^{(0)} +  \lb \begin{matrix}
		0 & \hat u\\
		u &0
	\end{matrix} \rb, \nonumber\\
&& V^{(2)}=   \lambda V^{(1)} + \lb \begin{matrix}
		-\hat u u & \partial_x\hat u\\
		-\partial_x u & u \hat u
	\end{matrix}\rb \nonumber\\
&& V^{(3)}=  \lambda V^{(2)} +  \lb \begin{matrix}
		\hat u\ \partial_xu - \partial_x\hat u\  u& -2 \hat u u \hat u +\partial_x^2 \hat u\\
		-2 u \hat u u + \partial_x^2 u  &u\  \partial_x \hat u - \partial_x u\ \hat u
	\end{matrix} \rb. \nonumber\\&&
\ldots
\end{eqnarray}
In general, the $V^{(n)}$ operator is  identified  as $V^{(n)} = \lambda V^{(n-1)} + w_0^{(n)}$. 
 Moreover, the recursion relations (\ref{recursion1})  lead to
\begin{equation}
w_k^{(n)} = w_{k-1}^{(n-1)}, ~~~~~w_{0}^{(n)} = (-1)^{n-1} w_{0}^{(1)}{\cal K}^{n-1},
\end{equation}
where the latter relations together with the constraints (\ref{constr2b}) suffice to provide 
$w_0^{(n)}$ at each order\footnote{Note that in the special case 
$\hat u = u$  the Lax pair $(U,\ V^{(3)})$ is the one of the mKdV model,  see also \cite{AKNS, Clark}.}.

We have thus far considered the simplest fundamental Darboux matrix, but nevertheless we were able to fully describe the dressing process.
Let us now briefly discuss the generic Darboux expressed as a formal $\lambda$-series expansion
\begin{equation}
{\mathrm G}(\lambda)= \lambda^m {\mathbb I}+ \sum_{k=0}^{m-1} \lambda^k g_k,
\end{equation}
where $g_{k}$ are $2  \times 2$ matrices to be identified.
We focus on the fundamental recursion relations arising from the $x$-part of the Darboux transform (\ref{fundam2}):
\begin{eqnarray}
&&  \lb \begin{matrix}
		0 & \hat u\\
		u &0
	\end{matrix} \rb= {1\over 2} \big [g_{m-1},\ \sigma \big ],  \label{c1}\\
&&\partial_x g_0 =   \lb \begin{matrix}
		0 & \hat u\\
		u &0
	\end{matrix} \rb\ g_0, ~~~~~ \partial_x g_k = {1\over 2} \big [ \sigma,\ g_{k-1} \big ]+  \lb \begin{matrix}
		0 & \hat u\\
		u &0
	\end{matrix} \rb\ g_k. \label{c2}
\end{eqnarray}
The infinite series expansion $m \to \infty$  is particularly interesting, especially via its connection to a 
Riccati equation of the type (\ref{Ric}) (see also for instance \cite{DoFiSk}). 


\section{Differential operators}
\noindent 
We have so far studied the Darboux-BT transform having chosen Lax pairs as well as Darboux transformations to be $d \times d$ matrices.
We now consider Lax pairs being differential operators and the Darboux transform expressed in terms of differential or integral operators.

Let us first recall the main ideas of the generalized Darboux-dressing scheme \cite{ZakharovShabat2}. Let
\begin{equation}
{\mathbb D}^{(n)}={\mathbb I}\partial_{t_n} + A^{(n)},~~~~{\mathbb L}^{(n)} = {\mathbb I} \partial_{t_n}+ {\mathbb A}^{(n)},
\end{equation}
where ${\mathbb I}$ is the $d \times d$ unit matrix and  $A^{(n)},\ {\mathbb A}^{(n)}$ are some 
``bare'' (free of fields) and ``dressed'' operators respectively. In general, they can be $d \times d$ matrices, matrix-differential or matrix-integral operators,
and are related via the generic Darboux-dressing transformation 
\begin{equation}
{\mathcal G}\ {\mathbb D}^{(n)}= {\mathbb L}^{(n)}\ {\mathcal G} \  \Rightarrow\  
\partial_{t_{n} }{\mathcal G} = {\mathcal G}  A^{(n)} - {\mathbb A}^{(n)} {\mathcal G}. \label{generalDarboux}
\end{equation}

Provided that $\big [{\mathbb D}^{(n)},\ {\mathbb D}^{(m)}\big ]=0$, the transformation (\ref{generalDarboux}) leads  to the generalized Zakharov-Shabat zero curvature relations, which define the integrable hierarchy
\begin{equation}
\partial_{t_n}{\mathbb A}^{(m)} - \partial_{t_m}{\mathbb A}^{(n)}+\big [{\mathbb A}^{(n)},\ {\mathbb A}^{(m)} \big]=0.
\end{equation}
In our setting here we choose to consider $A^{(n)},\  {\mathbb A}^{(n)}$  as differential operators, 
whereas the dressing transform ${\mathcal G}$ is chosen to be either a  differential or an integral one in the next subsection.
We employ our typical example i.e. the KdV equation, as our main paradigm to illustrate the general dressing process.

\subsection{Differential transforms: the KdV equation}
\noindent 
Before we describe the dressing process as a method of producing solutions as well as constructing the hierarchy we shall first 
derive the BT for the KdV equation using as the fundamental  Darboux-BT transformation the first order differential operator 
(see also for instance \cite{Kasman, Kodama}):
\begin{equation}
{\cal G} = \partial_x + K.
\end{equation}
Recall that the Lax pair for the KdV equation is given as
\begin{equation}
{\mathbb L}^{(2)} =- \partial_x^2 + u(x), ~~~~~{\mathbb L}^{(3)}= \partial_t - \alpha \partial_x^3 +a_1(x) \partial_x + a_0(x), \label{ll}
\end{equation}
where we define $a_1(x) = {3\over 2}\alpha u(x),\  a_0(x)= {3\over 4}\alpha \partial_x u(x) $.
The BT relations follow after implementing the transformation:
\begin{eqnarray}
&& {\cal G}\  \big ( -\partial_x^2 + \hat u \big ) = \big (  -\partial_x^2 +  u \big )\ {\cal G} \label{xx1}\\
&& {\cal G}\  \big ( \partial_t-\alpha\partial_x^3 + \hat a_1 \partial_x + \hat a_0 \big ) = \big ( \partial_t-\alpha\partial_x^3 +  
a_1 \partial_x + a_0  \big )\ {\cal G}.\label{tt1}
\end{eqnarray}
The solution of the latter equation leads to the BT relations for the KdV equation. Indeed, from (\ref{xx1}) we immediately obtain:
\begin{eqnarray}
&& u - \hat u= 2\partial_x K \nonumber\\
&& \partial_x^2 K = (u -\hat u) K - \partial_x \hat u. \label{tinda}
\end{eqnarray}
It is convenient to introduce the fields $w,\ \hat w$: $u = \partial_x w,\ \hat u = \partial_x \hat w$, then (\ref{tinda}) become
\begin{equation}
\partial_x \big (w +\hat w \big ) = {1\over 2} (w -\hat w)^2 -{\Lambda^2 \over 2}, \label{tind}
\end{equation}
where ${\Lambda ^2 \over 2}$ is some integration constant.
The $t$-part of the BT (\ref{tt1}) yields
\begin{equation}
\partial_t K - \alpha \partial_x^2 K + a_1 \partial_x K +\big (a_0 - \hat a_0 \big ) K-\partial_x \hat a_0 =0. \label{tdep}
\end{equation}
Taking also into account (\ref{tind}) we may re-express (\ref{tdep}) as (set $\alpha =-4$)
\begin{equation}
\partial_t \big (w - \hat w\big ) = 3\big (( \partial_x w)^2-(\partial_x\hat w)^2\big ) -\partial_x^3 \big ( w-\hat w\big ). \label{tdep1}
\end{equation}
Equations (\ref{tind}), (\ref{tdep1}) provide the B\"acklund transformation for the KdV equation (see e.g. \cite{Drazin} and references therein).

Let us now focus on the special case where $\hat w=0$, and derive the one-soliton solution for the KdV equation.  Eq. (\ref{tind}) then becomes
\begin{equation}
\partial_x w = { w^2\over 2} -{\Lambda^2 \over 2}, \label{basick}
\end{equation}
which can be easily solved, with the solutions being
\begin{equation}
w(x)  = \Lambda {1 +e^{\Lambda (x+ f(t))}\over 1- e^{\Lambda(x+f(t))}}. \label{1sol}
\end{equation}
Recall $u(x) = \partial_x w(x)$, which yields the one-soliton solution of the KdV equation as can be easily verified.

The time dependence of the solution above can be determined via the $t$-part of the BT (\ref{tdep1}) ($\hat w =0$):
\begin{equation}
\partial_t w = 3 (\partial_x w )^2- \partial_x^3 w. \label{basick1}
\end{equation}
But $w$ satisfies (\ref{basick}), which suggests
\begin{equation}
\partial_x^3 w= \partial_x \big (w\partial_x w \big ) = (\partial_x w)^2 + w^2\partial_x w. \label{basick2}
\end{equation}
Equations (\ref{basick1}), (\ref{basick2}) and (\ref{basick}) then yield
\begin{eqnarray}
\partial_t w = -\Lambda^2 \partial_x w,
\end{eqnarray}
and hence the function $f(t)$ appearing in (\ref{1sol}) becomes $f(t) = -\Lambda^2 t$.

We can also construct the two soliton solution via the soliton lattice construction and the use of the permutability theorem as we did for the sinh-Gordon model.
Indeed, consider (\ref{lattice}) for the KdV BT:
\begin{eqnarray}
&& \partial_x\big (w_i +w_0\big ) = {1\over 2} \big ( w_i -w_0\big )^2 - {\Lambda_i^2 \over 2}\nonumber\\
&&  \partial_x\big (w_{ij} +w_i \big ) = {1\over 2} \big ( w_{ij } - w_i\big )^2 - {\Lambda_j^2 \over 2}, ~~~~~i,\ j \in \{ 1,\  2\}. \label{BT1}
\end{eqnarray}
We now consider the special case where $w_0 =0$, and derive the two-soliton solution. The permutability requirement $w_{12} =w_{21}=w$ holds, and hence
\begin{eqnarray}
&&{w_2^2 \over 2} - {w_1^2 \over 2} - {\Lambda_2^2 \over 2} + {\Lambda_1^2 \over 2} = {1\over 2} \big (w -w_2\big )^2 -{1\over 2}\
\big (w -w_1\big )^2+{\Lambda_2^2 \over 2} - {\Lambda_1^2 \over 2}
\nonumber\\
&& \Rightarrow w ={\Lambda_1^2 -\Lambda_2^2\over w_1- w_2},
\end{eqnarray}
where $w_{1,2}$ are one soliton solutions (\ref{1sol}) with constants $\Lambda_1,\ \Lambda_2$ respectively. By taking the derivative of the  latter 
expression with respect to $x$  we obtain the 2-soliton solution for KdV.

\subsubsection*{Dressing}
\noindent
We briefly describe in what follows the dressing method  for the KdV equation utilizing the fundamental Darboux transform introduced above.
We introduce the ``bare''  differential operators:
\begin{equation}
{\mathbb D}^{(2)} = -\partial_x^2, ~~~~~{\mathbb D}^{(3)}= \partial_t - \alpha \partial_x^3. \label{dd2}
\end{equation}
and the ``dressed'' operators ${\mathbb L}^{(n)}$ are given by (\ref{ll}), where now $u,\ a_0$ and $a_1$ are to be determined via the dressing process.
Indeed, recall the fundamental Darboux-dressing transformation ${\cal G} = \partial_x + K$:
\begin{equation}
\big (\partial_x +K \big ){\mathbb D}^{(n)} = {\mathbb L}^{(n)}\big (\partial_x +K \big ),~~~~n\in \{2,\ 3\}.
\end{equation}
This leads to the following set of constraints from equation (\ref{dd2}) for $n=2$,
\begin{eqnarray}
&& u = 2\partial_x K \nonumber\\
&& \partial_x^2 K = u K , \label{tind2}
\end{eqnarray}
which provide a special case of  (\ref{tinda}) ($\hat u =0$), and yield the 1-soliton solution as already discussed.

The $t$-part of the dressing (\ref{dd2}) for $n=3$, leads to
\begin{eqnarray}
\partial_{t} K(x)- \alpha \partial_x^3 K(x)  + a_1(x) \partial_x K(x) +a_0(x) K(x) =0,
\end{eqnarray}
as well as to contributions proportional to $\partial_x^m,\  m\in \{ 1,\  2\}$, which essentially determine each one of the factors $a_k$ 
of the ${\mathbb L}^{(3)}$ operator in terms of the field $u$
\begin{eqnarray}
&& a_1(x) = {3 \alpha  \over2} u(x) \nonumber\\
&&a_0(x) = {3 \alpha  \over 4} \partial_x u(x),
\end{eqnarray}
where $u(x)$ is now given in (\ref{tind2}).

Although we have focused on the fundamental transform to describe the dressing we should note that the generic form of the Darboux 
can be expressed as an $M^{th}$ order differential operator
\begin{equation}
{\cal G} = \partial_x^M + \sum_{k=0}^{M-1}g_k(x)\partial_x^k,
\end{equation}
where the functions $g_{k}$ are to be derived via a set of constraints emerging from the dressing of the Lax pairs (see also e.g. \cite{DoFiSk}).
This is in analogy to the matrix case discussed in the previous section, where the general Darboux is expanded in a formal $\lambda$-power series.

\subsection{Integral transforms: the KdV equation}
\noindent 
Let us  introduce the necessary notation in the case where we choose the Darboux transform ${\cal G}$ to be an integral operator 
\cite{ZakharovShabat2, Ablo2},  and derive the Gelfand-Levitan-Marchenko (GLM) equation from the fundamental factorization condition. 

Let ${\mathcal G} = {\mathbb I} + {\mathcal K}^+$, and introduce the operators 
${\mathcal K}^{\pm},\ {\mathcal F}$ with integral representations 
($f$ is the test function):
\begin{eqnarray}
&& {\mathcal F}(f)(x) =\int_{\mathbb R} F(x,y) f(y)\ dy, \nonumber\\
&& {\mathcal K}^{\pm}(f)(x) =\int_{\mathbb R} K^{\pm} (x,y) f(y)\ dy, \label{kernels}
\end{eqnarray}
such that: $K^{+}(x, y) =0, ~~~x>y,$ and $K^-(x,y) =0, ~~~x<y.$
The operators ${\mathcal K}^{\pm},\ {\mathcal  F}$ are required to satisfy the factorization condition
\begin{equation}
({\mathbb  I} +{\cal  K}^+)\  ({\mathbb I} +{\cal F}) = {\mathbb I} +{\cal K}^-, \label{factorization}
\end{equation}
which leads to the fact that the kernel $K^+(x,y)$ satisfies the Gelfand-Levitan-Marchenko equation and 
$K^-(x, y)$ obeys an analogous integral equation:
\begin{eqnarray}
&& K^{+}(x,z) +F(x,z) + \int_x^{\infty}dy\ K^+(x,y)F(y,z)=0, ~~~~z>x, \nonumber\\
&& K^-(x, z) = F(x,z) + \int_x^{\infty}dy\ K^+(x,y)F(y,z), ~~~~~z<x.
\end{eqnarray}
$F(x, y)$ is the solution of the linear problem, i.e. invariance of the differential operators ${\mathbb D}^{(n)}$ under the action of the 
operator ${\cal F}$ is required:
\begin{equation}
{\cal F}\ {\mathbb D}^{(n)} = {\mathbb D}^{(n)}\ {\cal F}, ~~~~n\in \{2,\ 3 \}. \label{linearpde}
\end{equation}
Dependence on time $t$ is implied,  but omitted for now for brevity.

\subsubsection{Dressing}
\noindent 
We keep considering our main example i.e. the KdV equation in order to illustrate the dressing method using 
an integral operator as the Darboux-dressing transform.
The main aim is to identify the functions $u(x),\ a_1(x),\  a_0(x)$ appearing in the Lax pair (\ref{ll}) 
via the dressing  process as described below (see also \cite{ZakharovShabat2, Drazin}). 

We first perform the dressing for the time independent part of the Lax pair (i.e. ${\mathbb D}^{(2)} \to {\mathbb L}^{(2)}$,  (\ref{ll}), (\ref{dd2})):
\begin{eqnarray}
&& \int_x^{\infty}dy\ K(x,y)\big (-\partial_y^2\big )f(y) =\nonumber\\ &&u(x) f(x) -\partial_x^2\int_x^{\infty}dy\  K(x,y)f(y)  +\int_x^{\infty}dy\ 
u(x)K(x,y)f(y) \Rightarrow \nonumber\\ 
&& K(x,x) \partial_xf(x) -\partial_y K(x,y)|_{x=y} f(x) - \int_{x}^{\infty}dy\ \partial_y^2 K(x, y)f(y) = u(x) f(x) +\nonumber\\&& \partial_x\big ( K(x,x) f(x)\big )+
\partial_xK(x,y)|_{x=y}f(x) +\int_{x}^{\infty}dy\ \big (- \partial_x^2 + u(x)\big ) K(x,y)f(y). \nonumber\\
&& \label{gendress0}
\end{eqnarray}
From the expression above we immediately extract
\begin{eqnarray}
&& u(x)  = -2 \big (\partial_xK(x,y) + \partial_yK(x,y ) \big )|_{x=y}  \label{genfund}\\
&& -\partial_x^2 K(x, y) + \partial_y^2 K(x, y) + u(x) K(x, y) =0.
\end{eqnarray}

We next consider the dressing for the $t$-part of the Lax pair for generic $n^{th}$ order bare and dressed differential operators ($n\geq 3$) of the form:
\begin{equation}
{\mathbb D}^{(n)} = \partial_{t_n} - \partial_x^n, ~~~~~{\mathbb L}^{(n)} =  \partial_{t_n} - \partial_x^n + \sum_{k=1}^{n-1}a_k(x) \partial_x^k.
\end{equation}
Then the  basic dressing relation (\ref{generalDarboux}) is
\begin{eqnarray}
\int_x^{\infty}dy\ K(x,y)\big (\partial_{t_n}-\alpha \partial_y^n\big )f(y) &=& {\mathbb X}_x f(x) +\big  (\partial_{t_n}-\partial_x^n\big )\int_x^{\infty}dy\ 
K(x,y)f(y) \nonumber\\ &+&\int_x^{\infty}dy\ {\mathbb X}_xK(x,y)f(y),  \label{gendress}
\end{eqnarray}
where we define ${\mathbb X}_x = \sum_{k=0}^{n-1}a_k(x) \partial_x^k$. After repeated integrations by parts, and carefully taking into consideration 
the boundary terms by iteration, we conclude
\begin{equation}
 \partial_{t_n} K(x,y)  - \partial_x^n K(x,y )+ (-1)^n \partial_y^n K(x,y )+\sum_{k=0}^{n-1} a_k(x) \partial^k_x K(x,y) =0, \label{dress2}
\end{equation}
whereas the use of the boundary terms for each order $\partial_x^m f(x)$ provides the elements $a_k$. 
In the special case $n=3$, the solution of the above constraints leads as expected to: $a_2 =0,\ a_1 = {3\over 2 }u,\ a_0 = {3 \over 4} \partial_x u$, 
i.e. we recover the KdV Lax pair. The keen reader is encouraged to work out the dressing (\ref{gendress}), and obtain the generalized relation
 (\ref{dress2}) as well as the corresponding boundary contributions (see also e.g. \cite{Drazin, DoFiSk}). Note that we have assumed here $\alpha =1$, 
which corresponds to a  simple re-scaling of the time variable.

\subsubsection{Solving the Gelfand-Levitan-Marchenko equation}
\noindent 
In this section we are solving the Volterra-type integral GLM equation for given solutions of the linear problem (\ref{linearpde})
\begin{equation}
	K(x, z) + F(x + z) + \int_{x}^{\infty} K(x, y) F(y + z) d y = 0. \label{eq:GLM}
\end{equation}
We distinguish in what follows two cases  corresponding to the ``discrete'' and ``continuous'' solutions of the linear problem.

Let us first consider ``discrete''  solutions of the linear problem (\ref{linearpde}) of the generic form
\begin{equation}
	F(x,z,t ) = \sum_{n = 1}^{N} b_n \e{-\kappa_n( x+z) +\Lambda_n t}, \label{eq:nRNP_F}
\end{equation}
where the general dispersion relation immediately follows from (\ref{linearpde}): $\Lambda_{n} =-2\alpha  \kappa_n^3$.
This type of solution for the linear problem will yield the $N$-soliton solution of the KdV equation.
Given the form of the solution of the linear problem (\ref{eq:nRNP_F}) and the GLM equation we can express the kernel $K$ as 
\begin{equation}
K(x,z,t) = \sum_{n=1}^N L_n(x, t) e^{-\kappa_n x},
\end{equation}
where $L_n(x,t)$ are to be explicitly determined. Given that the kernel $K$ satisfies the GLM equation we obtain
\begin{align}
	0 &= L_n(x) + b_ \e{-\kappa_n x} + b_n \sum_{m = 1}^{N} L_m(x) \int_{x}^{\infty} \e{-(\kappa_n + \kappa_m) y} d y \nonumber \\
	&= L_n(x) + b_n \e{-\kappa_n x} + \sum_{m = 1}^{N} \frac{b_n}{\kappa_n + \kappa_m} L_m(x) \e{-(\kappa_n + \kappa_m) x}. \label{eq:nR1P_GLM}
\end{align}
The latter equation can be expressed in a compact form as
\begin{equation}
AL =- B \  \Rightarrow \ L = - A^{-1} B,
\end{equation}
where $ L $ and $ B $ are column $ N $-vectors with components $ L_n $ and $ b_n \e{-\kappa_n x} $ respectively, 
while $ A $ is an $ N \times N $ matrix with entries:
\begin{equation}
	 A_{nm} = \delta_{nm}+ \frac{b_n}{\kappa_n + \kappa_m} \e{-(\kappa_n + \kappa_m) x+\Lambda_n t}.
\label{eq:nRNP_A}
\end{equation}

An interesting observation is in order here. It is easy to check that
\begin{equation}
\partial_x A_{nm } = E_{n} B_{m}, \label{deriv}
\end{equation}
where $E_{n} = e^{-\kappa_n x}$. Also, recall that the kernel is expressed as (using also (\ref{deriv}))
\begin{equation}
K(x,x,t) = \sum_{n=1}^N L_n E_n = \sum_{n, m}  A_{nm}^{-1} B_m E_n = \sum_{n, m}A_{nm}^{-1} \partial_x A_{nm},
\end{equation}
which leads to
\begin{equation}
K(x, x ,t) = tr \big ( A^{-1} \partial_x A \big) = |A|^{-1} \partial_x|A| = \partial_x  (\ln {|A|}).
\end{equation}

We have however derived the field $u$ via dressing in the previous subsection (\ref{genfund}):
\begin{equation}
	u(x) =  -2\partial_x^2 \ln{|A|} , \label{eq:nRNP_uEq}
\end{equation}
and from the latter expression we readily derive the one soliton solution for the KdV equation
\begin{align}
	u(x) &= -4\kappa^2 \sech[2]{\kappa x + \Lambda t+ x_0},  \label{eq:nRNP_u}
\end{align}
where recall $\Lambda = -2\alpha \kappa^3$.

Let us now consider general solutions of the linear problem (\ref{linearpde}) expressed as 
continuous Fourier transforms
\begin{eqnarray}
&& F(x,z, t) = \int_{\mathbb R} dk\ b_k e^{ \Lambda_kt  +ik (x+ z)}.
\end{eqnarray}
As in the case of discrete solutions the dispersion relation is cubic:
\begin{eqnarray}
 \Lambda_k =-2 \alpha i k^3. \label{disp2}
\end{eqnarray}
Taking into consideration the solutions of the linear problem as well as the GLM equation we can express the kernel as
\begin{eqnarray}
K(x,z,t) =\int_{\mathbb R} dk\ {\mathbb L}(k, x,t)e^{ik z}.
\end{eqnarray}
Let us also define the quantities:
\begin{eqnarray}
 {\mathrm B}(k,x,t)= b_{k} e^{\Lambda_{k} t+ i  k x }, ~~~~~~{\mathbb P}(\hat k, k, x,t) = {e^{\Lambda_kt + i (k +\hat k)x} \over i(k+\hat k)}.
\end{eqnarray}
Provided that  the operator ${\mathbb I} - {\mathbb P}$ is  invertible  i.e. 
the Fredholm determinant is non-zero
($\det({\mathbb I} - {\mathbb P})  \neq 0$),
then ${\mathbb L}$ is explicitly identified  via the  integral equation
\begin{eqnarray}
&& \int d\hat k\  {\mathbb L}(\hat k, x,t)\Big  (\delta(\hat k, k) - {\mathbb P}(\hat k, k, x,t)\Big)= -{\mathrm B}(k,x, t).
\end{eqnarray}
The latter relation provides the formal series expansion for ${\mathbb L}$  i.e. 
the integral analogue of the matrix relation (\ref{eq:nR1P_GLM}) presented in the discrete case previously:
\begin{eqnarray}
{\mathbb L}(k) =-{\mathrm  B}(k) - \sum_{m=1}^{\infty}\int_{\mathbb R} dk_1 \ldots \int_{\mathbb R} dk_m\  
{\mathrm B}(k_1) {\mathbb P}(k_1, k_2)\ldots {\mathbb P}(k_m,k).
\end{eqnarray}

An interesting observation is in order here. Due to the cubic dispersion relation (\ref{disp2}) 
the solutions of the linear problem can be expressed in terms of Airy functions $\mbox{Ai}(x)$. Indeed, 
after expressing the coefficients $b_k$ in terms of the initial values 
of the solutions of the linear problem $F_0$ at $t=0$, via an inverse Fourier transform
\begin{equation}
b_k = {1\over 2\pi}\int_{\mathbb R} d\xi\  F_0(\xi)e^{-ik\xi},
\end{equation}
and recalling the definition of the Airy function
\begin{equation}
\mbox{Ai}(x) = {1\over \pi}\int_{0}^{\infty}dt\ \cos{{t^3\over 3} +xt},
\end{equation}
we obtain
\begin{equation}
F(x,z) ={1\over \nu} \int_{\mathbb R} d\xi\ \mbox{Ai}({x+z-\xi \over \nu})F_0(\xi),
\end{equation}
where we define $\nu = 2( 3t)^{1\over 3}$.

Distinct  choices of the initial profile $F_0$  give rise to different generic solutions, but an obvious choice of initial conditions is for instance, 
$F_0(\xi) =\delta(\xi)$.  Given that $F$ satisfies  the linear problem (\ref{linearpde}) ($j=3$), it is straightforward to see that  ${\mathbb F}(\zeta)= \nu F({x+z\over \nu}) $ 
satisfies as expected the Airy equation, (see also \cite{Ablo2}))
\begin{equation}
{\partial^2 {\mathbb F}(\zeta) \over \partial \zeta^2}- \zeta{\mathbb F}(\zeta) =0,
\end{equation}
with the parameter $\zeta$ defined as ${x+z\over \nu}$, and having also assumed that ${\mathbb F}(\zeta) \to 0$ when $\zeta \to \infty$. 
The kernel can be then expressed as a formal series expansion in terms of Airy functions. 

Similar interesting observations can be made for a second order linear differential operator. Indeed, in this case the solutions of the associated linear problem can be expressed in terms of the heat kernel, then via the so-called {\it Cole-Hopf transformation} another important non-linear PDE is obtained, 
called the viscous  Burgers equation (see for instance \cite{DoFiSk} and references therein).

\end{document}